\documentclass[10pt,prd,aps,floatfix]{revtex4}

\usepackage{graphicx}
\usepackage{amsmath}
\usepackage{amssymb}
\usepackage{longtable}

\newcommand{\be}{\begin{equation}}
\newcommand{\ee}{\end{equation}}
\newcommand{\bea}{\begin{eqnarray}}
\newcommand{\eea}{\end{eqnarray}}
\newcommand{\bi}{\begin{itemize}}
\newcommand{\ei}{\end{itemize}}
\newcommand{\nn}{\nonumber}

\begin{document}


\vspace*{-10mm}
\begin{flushright}
\normalsize
BNL-HET/06/6 \\
KANAZAWA-06-09 \\
KEK-CP-177
\end{flushright}

\title{
Vector form factor in $K_{l3}$ semileptonic decay 
with two flavors of dynamical domain-wall quarks
}

\author{
   Chris~Dawson$^{1}$, 
   Taku~Izubuchi$^{1,2}$,
   Takashi~Kaneko$^{3,4}$
   Shoichi~Sasaki$^{1,5}$, 
   and 
   Amarjit~Soni$^{6}$
}

\affiliation{
$^1$RIKEN-BNL Research Center, Brookhaven National Laboratory,
    Upton, NY 11973, USA \\
$^2$Institute for Theoretical Physics, Kanazawa University,
    Kanazawa, Ishikawa 920-1192, Japan \\
$^3$High Energy Accelerator Research Organization (KEK), 
    Tsukuba, Ibaraki 305-0801, Japan \\
$^4$Graduate University for Advanced Studies,
    Tsukuba, Ibaraki 305-0801, Japan \\
$^5$Department of Physics, University of Tokyo, Tokyo 113-0033, Japan \\
$^6$Physics Department, Brookhaven National Laboratory, 
    Upton, NY, 11973, USA
}

\date{\today}

\begin{abstract}

We calculate the vector form factor in $K \to \pi l \nu$ 
semileptonic decays at zero momentum transfer $f_+(0)$ 
from numerical simulations of two-flavor QCD on the lattice.
Our simulations are carried out on $16^3 \times 32$ 
at a lattice spacing of $a\!\simeq\!0.12$~fm 
using a combination of the DBW2 gauge and the domain-wall quark actions, 
which possesses excellent chiral symmetry even 
at finite lattice spacings.
The size of fifth dimension is set to $L_s\!=\!12$, which leads
to a residual quark mass of a few MeV.
Through a set of double ratios of correlation functions,
the form factor calculated on the lattice is accurately interpolated 
to zero momentum transfer, and then is extrapolated 
to the physical quark mass.
We obtain $f_+(0)\!=\!0.968(9)(6)$, where the first error is statistical
and the second is the systematic error due to the chiral extrapolation.
Previous estimates based on a phenomenological model and 
chiral perturbation theory are consistent with our result.
Combining with an average of the decay rate from recent experiments,
our estimate of $f_+(0)$ leads to the Cabibbo-Kobayashi-Maskawa (CKM)
matrix element $|V_{us}|\!=\!0.2245(27)$,
which is consistent with CKM unitarity.
These estimates of $f_+(0)$ and $|V_{us}|$ are subject to systematic 
uncertainties due to the finite lattice spacing and quenching of 
strange quarks, though 
nice consistency in $f_+(0)$ with previous lattice calculations 
suggests that these errors are not large.

\end{abstract}


\maketitle



\section{Introduction}
\label{sec:intro}


There has recently been rapid progress in the precise determination of 
the elements of the CKM matrix \cite{CKM} 
leading towards a stringent test of its unitarity.
Let us recall that such a test is a powerful method to search for 
new physics beyond the standard model.
In particular $d \to u$ and $s \to u$ semileptonic transitions provide
the most precise constraints on the size of the elements,
and hence CKM unitarity on the first row
\bea 
   |V_{ud}|^2 + |V_{us}|^2 + |V_{ub}|^2 = 1 - \delta
   \label{eqn:intro:CKM_unitarity}
\eea
can now be examined accurately~\cite{Vud_Vus,CKM2005}.


The values quoted in the 2004 PDG~\cite{PDG2004}
\bea
   |V_{ud}| = 0.9738(5), \hspace{5mm}
   |V_{us}| = 0.2200(26), \hspace{5mm}
   |V_{ub}| = (3.67 \pm 0.47) \times 10^{-3},
   \label{eqn:intro:CKM:PDG2004}
\eea
lead to 
\bea 
   \delta = 0.0033(15)
   \label{eqn:intro:delta:PDG2004},
\eea
which deviates from zero by two $\sigma$. We have to improve the accuracy on 
$\delta$ in order to confirm whether this deviation is a genuine signal of 
unitarity violation.
We note that $|V_{ub}|$ is so small that it can be safely neglected 
in this unitarity test.
Since about half the error of $\delta$ comes from the uncertainty in 
$|V_{ud}|$, and another half from $|V_{us}|$, we need a more precise 
determination of both of these two elements.
In the present paper, we focus on the determination of $|V_{us}|$.


We also note that $|V_{us}|$ gives the basic parameter 
$\lambda$
in the Wolfenstein parameterization of the CKM matrix \cite{Vus:W_param}.
A precise determination of $|V_{us}|$ is, therefore, 
important also for phenomenological studies of CP violation processes 
based on this parametrization.


So far, $|V_{us}|$ has been determined from several processes:
$K_{l3}$ decays \cite{PDG2004} which provide the value 
in Eq.~(\ref{eqn:intro:CKM:PDG2004}), 
hyperon $\beta$ decays \cite{Vus:hyperon},
$K_{\mu 2}$ and $\pi_{\mu 2}$ decays\cite{Vus:Kl2},
and hadronic $\tau$ decays\cite{Vus:tau_decay}.
At the moment, 
the $K_{l3}$ decays provide the most precise determination
among these, and its result has been quoted in the PDG.
We, therefore, try to determine $|V_{us}|$ through the $K_{l3}$ decays.


As will be explained in Sec.~\ref{sec:kl3},
$|V_{us}|$ can be determined through experimental determination of 
the decay rate $\Gamma$ and theoretical calculation of 
the vector form factor at zero momentum transfer $f_+(0)$.
The two $\sigma$ deviation from unitarity 
in Eq.~(\ref{eqn:intro:delta:PDG2004}) motivated recent measurements of 
$\Gamma$ \cite{kl3:exprt:br:E865,kl3:exprt:br:KTeV,kl3:exprt:br:NA48,kl3:exprt:br:KLOE,kl3:exprt:tau:KLOE}.
These experiments prefer a larger value for $|V_{us}|$,
which is consistent with unitarity.
However, a precise calculation of $f_+(0)$, say with an accuracy of 1\,\%,
is also indispensable in order to establish this consistency 
with unitarity.


A good theoretical control on $f_+(0)$ is provided by $SU(3)$ symmetry
and the Ademollo-Gatto theorem \cite{kl3:Ademollo-Gatto}.
The vector current conservation guarantees $f_+(0)\!=\!1$ at zero momentum 
transfer \cite{kl3:f2+f4:model:LR}, and then the Ademollo-Gatto theorem 
states that the $SU(3)$ breaking effects $f_+(0)\!-\!1$ start at second order 
in $(m_s\!-\!m_{ud})$, where $m_{ud}$ represents the averaged mass of 
up and down quarks.
This also indicates that the leading correction to $f_+(0)$ in chiral
perturbation theory (ChPT) does not contain the low-energy constants (LECs)
of the next-leading order chiral Lagrangian, and hence is practically
free of uncertainties.


In this paper, we calculate the vector form factor of 
the $K_{l3}$ decay at zero momentum transfer $f_+(0)$ non-perturbatively
from numerical simulations of lattice QCD with two degenerate flavors
of dynamical quarks, which are identified with up and down quarks.
Strange quarks are treated in the quenched approximation.
To make the best of use of the good theoretical control mentioned in the 
previous paragraph, we employ a combination of the DBW2 gauge action 
\cite{DBW2} and the domain-wall quark action \cite{DWF,DWF:sim}, 
which has excellent chiral symmetry even at finite lattice spacings.
We also employ the so-called double ratio method \cite{lat:dble_rat}
to improve the accuracy on the form factor.
Preliminary results of these calculations have been reported in 
Ref.\cite{kl3:fn:Nf2:RBC}.


This paper is organized as follows. 
We present a brief introduction of the $K_{l3}$ decays and status of 
experimental and theoretical studies on them in Sec.~\ref{sec:kl3}.
Our simulation method is introduced in Sec.~\ref{sec:sim_param}.
Section~\ref{sec:ff} is devoted to our determination of the form factor 
at finite momentum transfer from a double ratio of three-point functions.
We describe the interpolation of the form factor to zero momentum transfer
and the chiral extrapolation 
in Secs.~\ref{sec:q2_interp} and \ref{sec:chiral_fit}.
Section~\ref{sec:Vus} presents our estimate of $|V_{us}|$.
Our conclusions are given in Sec.~\ref{sec:conclusion}.


\section{$K_{l3}$ decays}
\label{sec:kl3}

\subsection{phenomenology of $K_{l3}$ decays}
\label{sucsec:kl3}


The $K_{l3}$ decays are $K$ to $\pi$ semileptonic decay channels
\bea
   K_{l3}^0 &:& K^0 \to \pi^- \l^+ \nu_l,
   \label{eqn:kl3:kl30}
   \\    
   K_{l3}^+ &:& K^+  \to \pi^0 \l^+ \nu_l,
   \label{eqn:kl3:kl3+}
\eea
where $l$ represents the electron or muon.
In the following, we mainly consider the neutral kaon decay $K_{l3}^0$.
A simplification in theoretical studies of these decays
is 
that the matrix element of the axial current 
vanishes due to the parity symmetry. 
Therefore, the decay amplitude contains only the matrix element
of the vector current $V_\mu = \bar{s} \gamma_\mu u$ which can be 
expressed in terms of form factors
\bea
   \left\langle 
      \pi(p^{\prime}) \left| V_\mu \right| K(p)
   \right\rangle
   & = & 
   \left( p_{\mu} + p^{\prime}_{\mu} \right) \, f_+(q^2) 
  +\left( p_{\mu} - p^{\prime}_{\mu} \right) \, f_-(q^2),
   \label{eqn:kl3:f+-}
\eea
where $q\!=\!p-p^\prime$ represents the momentum transfer.


In literature, the so-called scalar form factor 
\bea 
   f_0(q^2)
   & = &
   f_+(q^2) + \frac{q^2}{M_K^2 - M_{\pi}^2} f_-(q^2),
   \label{eqn:kl3:f0}
\eea
and 
\bea
   \xi(q^2) 
   & = &
   \frac{f_-(q^2)}{f_+(q^2)}
   \label{eqn:kl3:xi}
\eea
are often used instead of $f_-(q^2)$. 
In particular, $f_0(q^2)$ is a useful quantity in lattice calculations,
since i) it equals to $f_+(0)$, which appears in the expression 
of the decay rate (see Eq.~(\ref{eqn:kl3:decay_rate}) below), 
at zero momentum transfer $q^2\!=\!0$, 
ii) it can be precisely calculated from the matrix element with
kaon and pion momenta equal to zero 
\bea
   \left\langle 
      \pi(0) \left| V_4 \right| K(0)
   \right\rangle
   & = & 
   (M_K+M_\pi)\, f_0(q_{\rm max}^2),
   \label{eqn:kl3:ME2f0}
\eea
where $q_{\rm max}^2\!=\!(M_K-M_\pi)^2$.


The rate of the $K_{l3}$ decays is given by \cite{kl3:f2+f4:model:LR}
\bea
   \Gamma
   & = &
   \frac{G_\mu^2}{192\pi^3} M_K^5 C^2 \, I \,
   |V_{us}|^2 \, |f_+(0)|^2 \, S_{\rm ew}(1+\delta_{\rm em}),
   \label{eqn:kl3:decay_rate}
\eea
where $I$ and $S_{\rm ew}(1+\delta_{\rm em})$ represent the phase space 
integral and radiative corrections, respectively.
The Clebsch-Gordan coefficient $C^2=1(1/2)$ for the neutral
(charged) kaon decay is written explicitly in the above expression
so that $f_+(0)$ for both decay channels equals to unity
in the $SU(3)$ symmetric limit.


The phase space integral $I$ is generally defined as 
\cite{kl3:f2+f4:model:LR}
\bea
   I 
   & = &
   \frac{1}{M_K^8}
   \int d(q^2) \, \lambda_I^{3/2}
                  \left( 1 + \frac{M_l^2}{2\,q^2} \right)
                  \left( 1 - \frac{M_l^2}{q^2}  \right)^2
   \nn \\
   &   &
   \hspace{20mm}
   \times      \left\{ \frac{f_+(q^2)}{f_+(q_0^2)}
                      +\frac{3\,M_l^2\, (M_K^2-M_\pi^2)^2}
                            {(2\,q^2 + M_l^2)\lambda_I}
                       \frac{f_0(q^2)}{f_+(q_0^2)}
               \right\},
   \label{eqn:kl3:I} \\
   \lambda_I
   & = &
   q^4 + M_K^4 + M_{\pi}^4 
       - 2\,q^2\,M_K^2 -2\,q^2\,M_{\pi}^2- 2\,M_{K}^2\,M_{\pi}^2,
\eea
where $q_0$ is a reference value of the momentum transfer.
If we take $q_0 \! \neq \! 0$, $f_+(0)$ in Eq.~(\ref{eqn:kl3:decay_rate}) 
has to be replaced by $f_+(q_0)$.
As in Eq.~(\ref{eqn:kl3:decay_rate}), 
$q_0$ is usually set to 0 so that $I$ depends on $f_{+,0}(q^2)$
only through small coefficients
$\lambda_+^{(1)}$, $\lambda_+^{(2)}$ and $\lambda_0^{(1)}$,
which parametrize the $q^2$ dependence of $f_{+,0}(q^2)$ 
\bea
   f_{+}(q^2)
   & = &
   f_+(0)\,\left(1+\lambda_+^{(1)}\,q^2
                  +\lambda_+^{(2)}\,q^4
               \right), 
   \label{eqn:kl3:q2_dep:f_+}
   \\
   f_0(q^2)
   & = &
   f_0(0)\,\left(1 + \lambda_0^{(1)}\,q^2 \right), 
   \label{eqn:q2_interp:form:lin}
\eea
where we include the quadratic term suggested by the KTeV experiment to
$f_+(q^2)$ \cite{kl3:exprt:lambda:KTeV}.
%
%
From recent experimental measurements of these coefficients 
\cite{kl3:exprt:lambda:KTeV,kl3:exprt:lambda:NA48,kl3:exprt:lambda:ISTRA}, 
the current estimate for $I$ is $\sim$ 0.154 (0.159) 
for $K^0_{e3}$ ($K^+_{e3}$) and 0.102 for $K^0_{\mu 3}$
with an accuracy of around 1\% \cite{Vud_Vus,CKM2005},
where a dominant error comes from the choice of the parametrization form of 
the $q^2$ dependence of $f_+(q^2)$.
The choice of the reference scale $q_0\!=\!0$ forces us to study 
the $q^2$ dependence of the form factor and take 
the limit of $q^2\!=\!0$ as in Sec.~\ref{sec:q2_interp}.


The radiative corrections split into the short-distance electroweak piece 
$S_{\rm ew}$ and the long-distance electromagnetic piece $(1+\delta_{\rm em})$.
The former is precisely determined as $S_{\rm ew}\!=\!1.022$ \cite{kl3:Sew}.
Chiral perturbation theory including the electromagnetic interaction
\cite{kl3:delta_em:ChPT} and a phenomenological model \cite{kl3:delta_em:model}
reveal that the latter is small correction and its uncertainty
leads to $\lesssim 1\%$ error to $\Gamma$.


Recently several new experimental determination of $\Gamma$ have been 
performed to clarify the origin of the two $\sigma$ deviation from unitarity 
in Eq.(\ref{eqn:intro:delta:PDG2004})
\cite{kl3:exprt:br:E865,kl3:exprt:br:KTeV,kl3:exprt:br:NA48,kl3:exprt:br:KLOE,kl3:exprt:tau:KLOE}.
In Ref.\cite{CKM2005}, $|V_{us}\, f_+(0)|\!=\!0.2173$ is obtained 
from the new measurements of $\Gamma$ and recent estimates of $I$
and $\delta_{\rm em}$ through Eq.~(\ref{eqn:kl3:decay_rate}).
This is about 3\% larger than 0.2114 corresponding to 
$|V_{us}|$ in Eq.~(\ref{eqn:intro:CKM:PDG2004}), and may lead
to a good consistency with CKM unitrarity.
However, we have to determine $f_+(0)$ with an accuracy of 1\% 
in order to make a definite conclusion on this unitarity test.

\subsection{previous theoretical studies of $f_+(0)$}
\label{sucsec:prev}


In previous theoretical studies,
$f_+(0)$ is considered in the following ChPT expansion 
\bea
   f_+(0) = 1 + f_2 + f_4 + O(p^6),
   \label{eqn:kl3:f+:expand}
\eea 
where $f_{2n}$ is the $O(M_{\pi,K,\eta}^{2n}) \!=\! O(m_q^n)$ correction
to $f_+(0)$.
We note that the leading term is unity, because 
$f_+(0)$ becomes the Clebsch-Gordan coefficient in the $SU(3)$ symmetric limit
thanks to the vector current conservation, and 
we explicitly factor it out Eq.~(\ref{eqn:kl3:decay_rate}).


The Ademollo-Gatto theorem\cite{kl3:Ademollo-Gatto} states that 
$SU(3)$ breaking effects are second order in $(m_s-m_{ud})$.
From previous theoretical studies, it turned out that 
the $SU(3)$ breaking effects $f_+(0)\!-\!1$ are order of 3\,--\,5\%.
Therefore, we can achieve 1\% accuracy on $f_+(0)$ by calculating 
the $SU(3)$ breaking effects with an accuracy of 20\,--\,30\%,
which is not prohibitively challenging.


In addition, the Ademollo-Gatto theorem guarantees that 
the leading correction $f_2$ does not contain any poorly-known LECs, 
which are associated with analytic terms 
from the $O(p^4)$ chiral Lagrangian.
For example, its ChPT formula for the $K_{l3}^0$ decay 
is given by \cite{kl3:f2:ChPT:GL}
\bea
   f_2
   & = &
   H_{K^0 \pi} + \frac{1}{2} H_{K^+ \pi}
               + \frac{3}{2} H_{K^+ \eta}
               + \epsilon \sqrt{3} (H_{K \pi} - H_{K \eta}),
   \label{eqn:kl3:f2}
  \\
   H_{PQ}
   & = &
   - \frac{1}{64\pi^2f_{\pi}^2}
   \left\{ M_P^2 + M_Q^2 
         +\frac{2 M_P^2 M_Q^2}{M_P^2-M_Q^2} 
          \ln \left[ \frac{M_Q^2}{M_P^2} \right]
   \right\},
   \label{eqn:kl3:HPQ}
\eea
where $\epsilon \! = \! (\sqrt{3}/4)(m_d-m_u)/(m_s-m_{ud})$.


However, the next leading correction $f_4$ 
contains LECs in $O(p^4)$ and $O(p^6)$ chiral Lagrangians
\cite{kl3:f4:ChPT:BCE,kl3:f4:ChPT:PS,kl3:f4:ChPT:BT,kl3:f4:ChPT:JOP,kl3:f4:ChPT:CEEKPP}.
Therefore, $f_4$ is difficult to determine only from ChPT,
and hence a phenomenological estimate $f_4\!=\!-0.016(8)$ 
by Leutwyler and Roos \cite{kl3:f2+f4:model:LR}
has been employed in previous determinations of $|V_{us}|$.


Clearly, it is desirable to calculate $f_+(0)$ non-perturbatively.
This background led to the first lattice study in quenched QCD 
\cite{kl3:fn:Nf0:italy}. 
They demonstrated that lattice calculations can achieve the 1\% accuracy for 
$f_+(0)$ by using a set of the so-called double ratios of correlation 
functions, and by making good use of the ChPT formula for $f_2$,
namely Eq.(\ref{eqn:kl3:f2}), in the chiral extrapolation of their
lattice data.
They employed the non-perturbatively $O(a)$-improved Wilson quark action
and obtained $f_+(0)\!=\!0.960(9)$,
which is consistent with the Leutwyler-Roos's estimate.



The calculation was extended to two-flavor QCD 
with the $O(a)$-improved Wilson quark action \cite{kl3:fn:Nf2:JLQCD}
and to three-flavor QCD with an improved Kogut-Susskind (KS) quark action
\cite{kl3:fn:Nf3:FNAL}.
While the $q^2$ dependence of $f_+(q^2)$ has not been investigated 
in the latter study, their estimates of $f_+(0)$ are consistent
with that in quenched QCD.



It is also worth while to note that the so-called twisted boundary
condition \cite{kl3:tbc}, which enables us to explore small $q^2$ region,
has been tested in quenched QCD \cite{kl3:fn:Nf0:tbc}.


\section{Simulation method}
\label{sec:sim_param}


In this study, we calculate the kaon form factor $f_+(0)$ 
by numerical simulations of lattice QCD with two-flavors of dynamical quarks, 
which are identified with up and down quarks.
We note that the isospin breaking effects in Eq.~(\ref{eqn:kl3:f2})
is proportional to $\epsilon \! \approx \! 0.01$, and $f_2$ itself is of order
2\,--\,3\% shift, so the correction due to isospin breaking
is well below our target accuracy on $f_+(0)$ of 1\%.
Strange quarks are treated in the quenched approximation.
We employ the domain-wall quark action \cite{DWF}, 
which has the following advantages over the conventional  
Wilson- and KS-type actions.


First, it possesses chiral symmetry even at finite lattice spacings 
in the limit of $L_s \to \infty$, while
the conventional Wilson- and KS-type fermions break 
the chiral symmetry explicitly. 
The chiral behavior of physical quantities may be distorted 
with the conventional fermions. Rigorously speaking, 
we have to take account of effects of the explicit symmetry breaking 
in the chiral extrapolation of the quantities obtained with 
the conventional fermions at finite lattice spacings \cite{WChPT,SChPT}.
In contrast, chiral extrapolations with domain-wall quarks
are fairly straightforward and simple as they need use
essentially continuum ChPT~\cite{RBC_bilinear}.
This is particularly important in this study of the $K_{l3}$ from factor, 
since we can safely subtract the leading correction $f_2$ 
from our lattice data of $f_+(0)$ 
by using the ChPT formula Eq.(\ref{eqn:kl3:f2})
before the chiral extrapolation so that 
systematic uncertainties due to the extrapolation influence
the final result only through the small higher order corrections.


Another advantage of the use of domain-wall quark action is that 
it is automatically $O(a)$-improved. 
Unlike the Wilson-type fermions, 
the non-perturbative tuning of improvement coefficients for the vector current
is not necessary to remove possibly large $O(a)$ effects \cite{Oa_effects}.
We also note that
the leading $O(a^2)$ scaling violation in physical quantities 
such as the kaon $B$ parameter is not large at $a^{-1}\simeq2$~GeV
\cite{BK:Nf0QCD:CP-PACS,BK:Nf0QCD:RBC}.


These virtues of domain-wall fermions are lost if $L_s$ is not sufficiently 
large.
Since the CPU cost is proportional to $L_s$, 
it is limited to values around 10\,--\,20 in practical unquenched 
simulations.
In order to improve chiral properties of the form factors with $L_s$
fixed, we employ the DBW2 gauge action \cite{DBW2} with which
the light hadron spectrum and the kaon $B$ parameter 
show better chiral properties than with the conventional plaquette 
action\cite{Spectrum:Nf0QCD:RBC,Spectrum+BK:Nf2QCD:RBC}.


We use gauge ensembles generated 
on a $L^3 \times T \!=\! 16^3 \times 32$ lattice at $\beta=0.80$,
as discussed in Ref.\cite{Spectrum+BK:Nf2QCD:RBC}.
The lattice spacing is $a \simeq 0.12$~fm and 
the physical spatial size is $La \simeq 1.9$~fm.
We set the domain-wall height to $M_5\!=\!1.8$ and 
the fifth-dimensional length to $L_s\!=\!12$.
The resulting residual quark mass is a few MeV.
We simulate sea quark masses $m_{ud,\rm sea}\!=\!0.02$, 0.03, and 0.04
in the range of $m_{s,\rm phys}/2 \lesssim m_{ud,\rm sea} \! \lesssim \! m_{s,\rm phys}$,
where $m_{s,\rm phys}$ represents the physical strange quark mass.
Our statistics are 94 configurations separated by 50 HMC trajectories
at each sea quark mass.
We refer to Ref.~\cite{Spectrum+BK:Nf2QCD:RBC} for further details 
on the configuration generation.


On these gauge ensembles,
we calculate two and three point functions
\bea
   C^{P}(t;{\bf p})
   & = &
   \sum_{\bf x}
   \left\langle 
      {\mathcal O}_{P,\rm snk}({\bf x},t+t_0) \,
      {\mathcal O}_{P,\rm src}^{\dagger}({\bf 0},t_0)
   \right\rangle
   \, 
   e^{-i{\bf p}{\bf x}},
   \nn \\
   & \xrightarrow[t \to \infty]{} &
   \frac{Z_{P,\rm src}^*\, Z_{P,\rm snk}}
        {2\, E_P({\bf p})}\,
   e^{-E_P({\bf p})\,t}
	  \label{eqn:sim:2pt}
	  \\
   C_{\mu}^{PQ}(t,t^{\prime};{\bf p},{\bf p}^{\prime})
   & = &
   \sum_{{\bf x},\, {\bf x}^{\prime}}
   \left\langle 
      {\mathcal O}_{Q,\rm snk}({\bf x}^{\prime},t^{\prime}+t_0) \,
      V_{\mu}({\bf x},t+t_0) \, 
      {\mathcal O}_{P,\rm src}^{\dagger}({\bf 0},t_0)
   \right\rangle
   \, 
   e^{-i{\bf p}^{\prime}({\bf x}^{\prime}-{\bf x})}
   e^{-i{\bf p}{\bf x}},
   \nn \\
   & \xrightarrow[t,(t^{\prime}-t) \to \infty]{} &
   \frac{Z_{P,\rm src}^*\, Z_{Q,\rm snk}}
   {4\, E_P({\bf p}) E_Q({\bf p}^{\prime})\, Z_V}
   \langle Q(p^{\prime}) | V_\mu^{\rm (R)} |P(p)\rangle\,
   \nn \\
   &   &
   \hspace{40mm} 
   \times 
   e^{-E_{P}({\bf p})\,t -E_{Q}({\bf p}^{\prime})\,(t^\prime-t)}
	  \label{eqn:sim:3pt}
\eea
where $P$ and $Q$ denote $K$ or $\pi$ meson.
The sink (source) operator for the meson $P$ is represented by 
${\mathcal O}_{P,\rm snk(src)}^{(\dagger)}$, 
and its overlap to the physical meson state is given 
by $Z_{P,\rm snk(src)}\!=\!\langle 0 | O_{P, \rm snk(src)} | P \rangle$.
We denote the energy of meson $P$ with a spatial momentum $\bf p$ 
by $E_{P}({\bf p})$.
The renormalized vector current with the renormalization factor $Z_V$
is represented by $V_\mu^{(R)}$.

In our preliminary study \cite{kl3:fn:Nf2:RBC},
an exponential smeared operator
\bea
   && 
   \sum_{\bf r}\phi(|{\bf r}|)\, \bar{q}({\bf x}) \, \gamma_5 
                              \, q({\bf x}+{\bf r}),  
   \hspace{5mm} \phi(|{\bf r}|) = A\,\exp\left[ -B\,|{\bf r}|\right]
   \label{eqn:sim:smr}
\eea
with $A\!=\!1.2$ and $B\!=\!0.1$ was used for the initial meson.
We observe that the correlators $C^{P}(t;{\bf p})$ and 
$C_{\mu}^{PQ}(t,t^{\prime};{\bf p},{\bf p}^{\prime})$ with non-zero ${\bf p}$
show poor signals with this choice of the smearing function. This may 
suggest that this operator is too close to the wall source,
and the correlators have small overlap with meson states with non-zero ${\bf p}$.
In this study, therefore, we use more localized operators 
with $B\!=\!0.5$, 0.6, and 0.7 at $m_{ud}\!=\!0.02$, 0.03, and 0.04,
respectively.
We take a single choice of $t_0\!=\!4$ for both of $C^{P}(t;{\bf p})$ 
$C_{\mu}^{PQ}(t,t^{\prime};{\bf p},{\bf p}^{\prime})$.
The local operator, which is combined with the sequential source 
method \cite{SSM} for $C_{\mu}^{PQ}(t,t^{\prime};{\bf p},{\bf p}^{\prime})$,
is used for the sink meson operator.


As in Ref.\cite{Spectrum+BK:Nf2QCD:RBC},
we calculate the quark propagator 
with each of the periodic and anti-periodic boundary conditions 
in temporal direction for quarks.
The correlation functions $C^{P}$ and $C_\mu^{PQ}$ are constructed 
by using the averaged quark propagator over the boundary conditions.
This procedure cancels effects of valence quarks 
wrapping the lattice in the temporal direction by odd number of times,
and enable us to take the time slice for the sink operator 
$(t^\prime\!+\!t_0)$ for $C_\mu^{PQ}$ larger than $T/2$.
In this study, we fix $t^\prime$ to 24 for all combinations of 
sea and valence quark masses.


In our measurement of $C^{P}$ and $C_\mu^{PQ}$, we fix the 
valence $ud$ quark mass equal to the sea quark mass, 
and take four strange quark masses $m_s\!=\!0.02$, 0.03, 0.04, and 0.05, 
which are roughly in a range of $[m_{s,\rm phys}/2,5\,m_{s,\rm phys}/4]$. 
For the meson momentum,
we take all possible configurations $\bf p$ with 
$|{\bf p}|^2\!=\!0,1$, and 2 for the initial meson ``$P$'' 
in $C^P$ and $C_\mu^{PQ}$. 
For notational simplicity, we use the momentum $p_k$ 
in units of $2\pi/L_k$ throughout this paper.
Two configurations ${\bf p}\!=\!(0,0,0)$ and $(-1,0,0)$ are used 
for the final meson ``$Q$'' in $C_\mu^{PQ}$.


\section{Scalar form factor at $q_{\rm max}^2$}
\label{sec:ff}

\subsection{Meson masses}
\label{subsec:ff:mass}

\begin{figure}[b]
\vspace{2mm}
\begin{center}
\includegraphics[width=0.5\linewidth,clip]{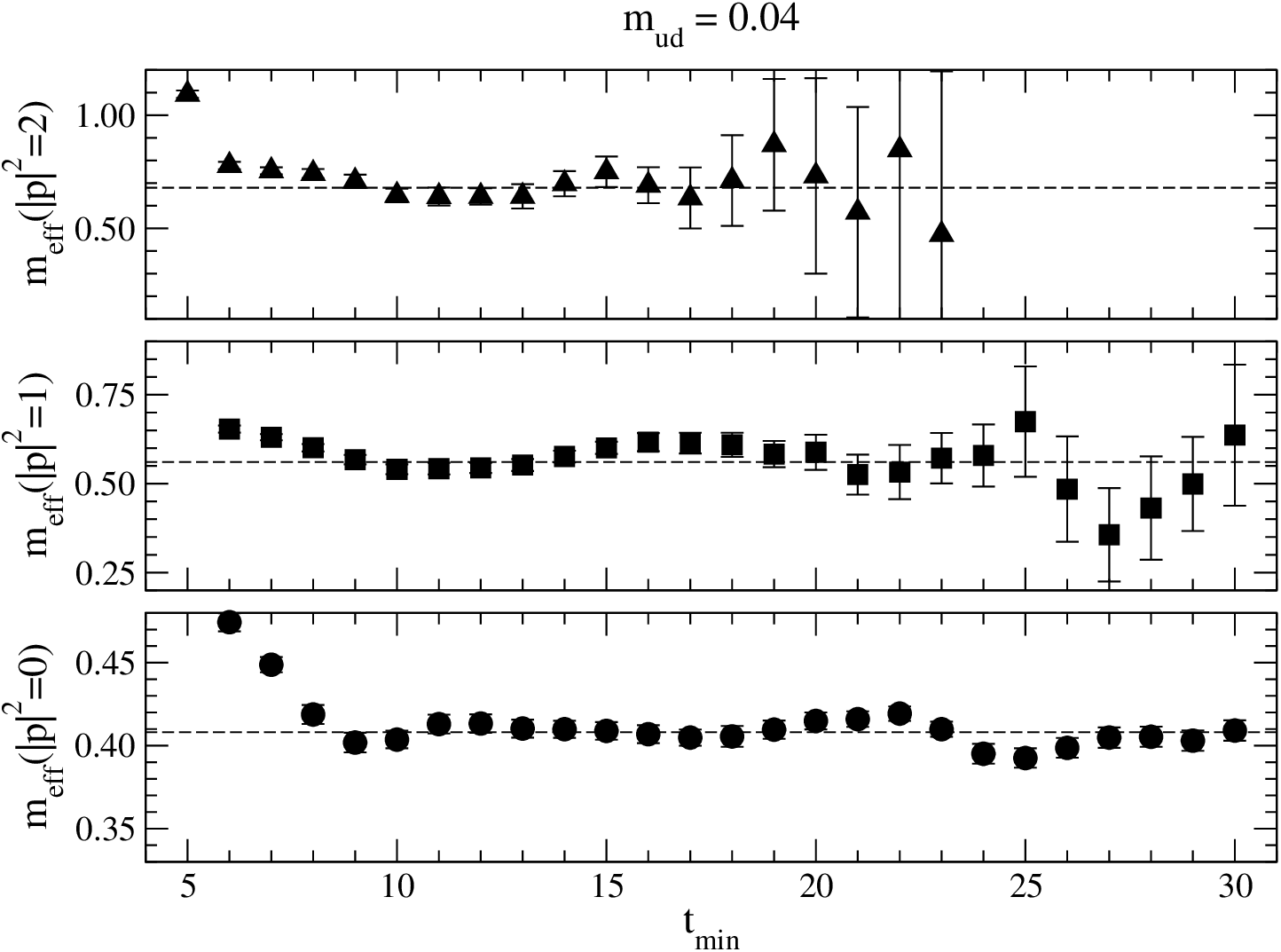}
\vspace{8mm}

\includegraphics[width=0.5\linewidth,clip]{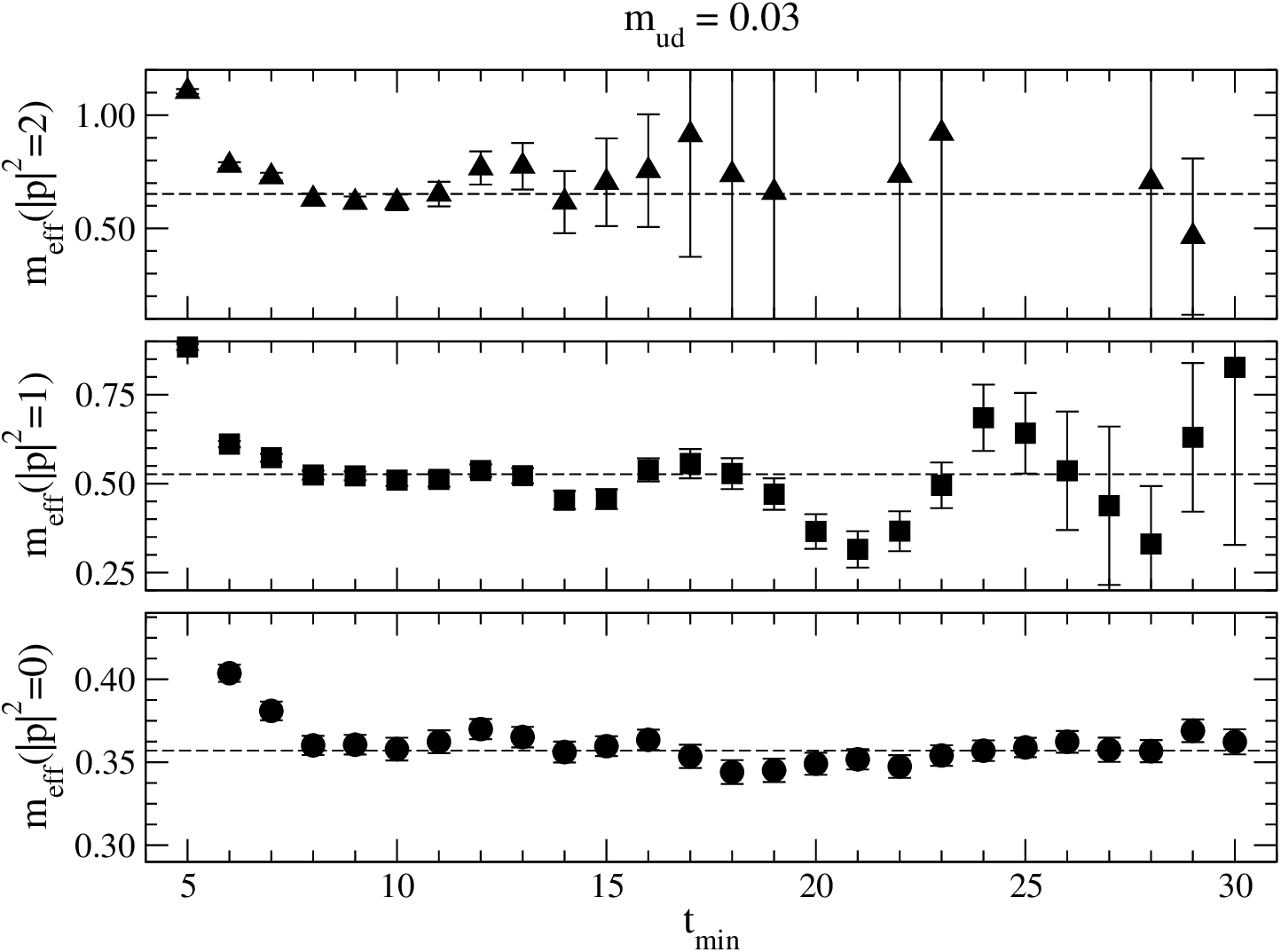}
\vspace{8mm}

\includegraphics[width=0.5\linewidth,clip]{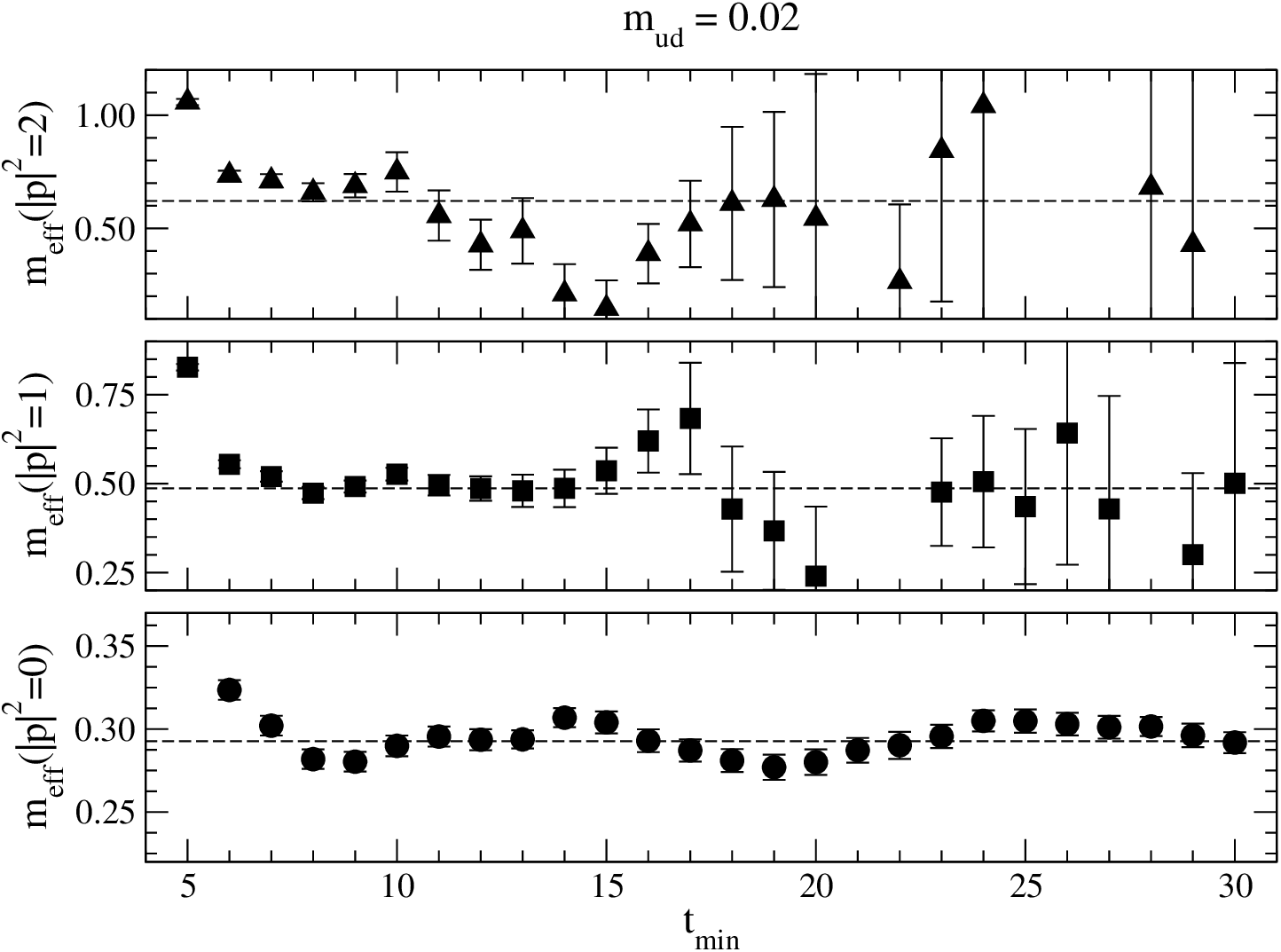}

\end{center}
\vspace{0mm}
\caption{
   Effective mass plots for pion at $m_{ud}\!=\!0.04$ (top figure),
   0.03 (middle figure), and 0.02 (bottom figure).
   Lines for data with zero meson momentum represent the fitted mass, 
   while those for larger momentum are an estimation 
   from the lattice dispersion relation Eq.~(\ref{eqn:ff:mass:ldr}).
}
\label{fig:ff:mass:em}
\end{figure}

In extraction of the form factors from $C_\mu^{PQ}$, we need 
precise knowledge of the pion and kaon masses and 
their energies with finite momenta,
which appear in Eq.~(\ref{eqn:sim:3pt}).
In Fig.~\ref{fig:ff:mass:em}, we plot the effective mass for pions 
calculated from $C^{\pi}$.
We observe a clear and long plateau in the effective mass, 
and hence the (lattice) meson masses summarized in Table~\ref{tbl:ff:mass:mass}
are determined with an accuracy of $\lesssim$~1\%.
We note that masses presented in Ref.\cite{Spectrum+BK:Nf2QCD:RBC},
in which the local and wall sources are employed, 
are consistent with ours within statistical error.

\begin{figure}[b]
\vspace{2mm}
\begin{center}

\includegraphics[width=0.5\linewidth,clip]{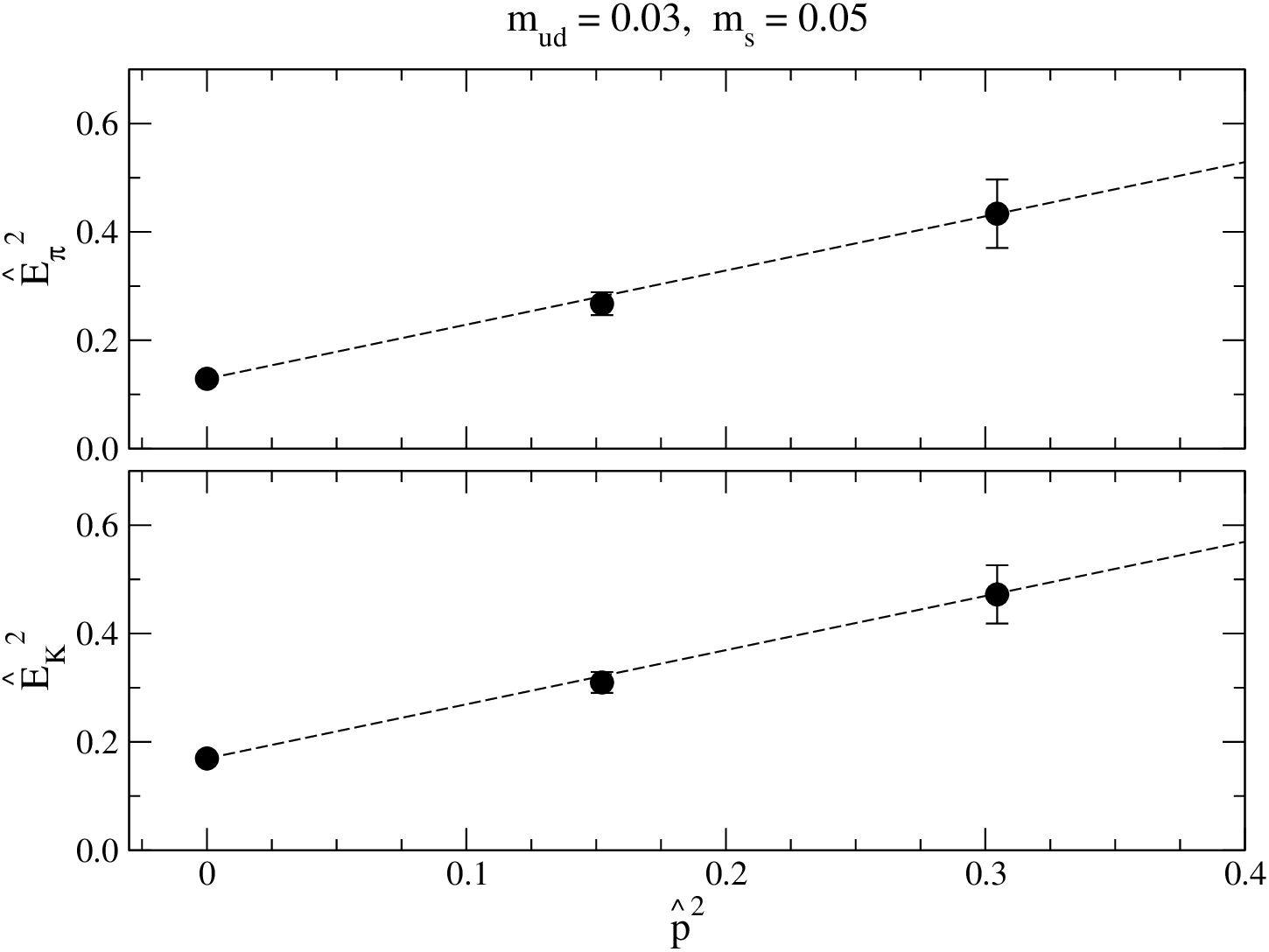}

\end{center}
\vspace{0mm}
\caption{
   Comparison of fitted energies (circles) with lattice dispersion relation
   Eq.~(\ref{eqn:ff:mass:ldr}) at $(m_{ud},m_s)\!=\!(0.03,0.05)$.
   Top and bottom panels show pion and kaon energies, respectively.
}
\label{fig:ff:mass:ldr}
\end{figure}

For the meson energies with non-zero momenta,
we use an estimation from the fitted mass $M$ and the lattice dispersion
relation
\bea
  \hat{E}({\bf p})^2
  & = &
  M^2 + \hat{\bf p}^2, \hspace{5mm}
  \hat{E}({\bf p}) =
  2\, \mbox{sinh}\left[\frac{E({\bf p})}{2}\right], \hspace{5mm}
  p_k = 2 \sin \left[\frac{p_k}{2}\right],
  \label{eqn:ff:mass:ldr}
\eea
instead of the fitted energy to $C^{P}$, since its statistical error 
rapidly increases as the size of the meson momentum increases.
We observe that i) the fitted energy shows a good consistency 
with Eq.~(\ref{eqn:ff:mass:ldr}) as in Fig.~\ref{fig:ff:mass:ldr};
ii) the final result for $f_+(0)$ at the physical quark mass 
does not change significantly if we use the dispersion relation
in the continuum limit instead of Eq.~(\ref{eqn:ff:mass:ldr}).

\subsection{$f_0(q_{\rm max}^2)$}
\label{subsec:ff:f0}


We consider the double ratio which was originally proposed 
in Ref.\cite{lat:dble_rat} for $B$ meson decays
\bea
   R(t)
   & = &
   \frac{C_{4}^{K\pi}(t,t^{\prime};{\bf 0},{\bf 0}) \, 
         C_{4}^{\pi K}(t,t^{\prime};{\bf 0},{\bf 0})}
        {C_{4}^{KK}(t,t^{\prime};{\bf 0},{\bf 0}) \, 
         C_{4}^{\pi\pi}(t,t^{\prime};{\bf 0},{\bf 0})},
   \label{eqn:ff:f0:R14}
\eea
where we fix $t^\prime$ to 24 as mentioned in Sec.~\ref{sec:sim_param}, 
and $t^\prime$ dependence of $R$ is ignored in the following.
All of $Z_V$, $Z_{P,\rm src}$, $Z_{Q, \rm snk}$ and the exponential damping 
factor in Eq.~(\ref{eqn:sim:3pt}) are exactly canceled in $R(t)$.
As a result, $R(t)$ contains only meson matrix elements with 
zero momentum, and gives $f_0(q_{\rm max}^2)$
\bea
   R(t)
   & \xrightarrow[t,(t^{\prime}-t) \to \infty]{} &
   \frac{\langle \pi | V_4^{\rm (R)} | K   \rangle\,
         \langle K   | V_4^{\rm (R)} | \pi \rangle}
        {\langle K   | V_4^{\rm (R)} | K   \rangle\,
         \langle \pi | V_4^{\rm (R)} | \pi \rangle}
   \hspace{2mm} 
   = 
   \hspace{2mm} 
   \frac{(M_K+M_\pi)^2}{4 M_K M_\pi} |f_0(q_{\rm max}^2)|^2.
   \label{eqn:ff:f0:R14tof0}
\eea
We note that $R(t)$ is exactly equal to unity in the $SU(3)$ symmetric limit
and hence it is a useful quantity to measure $SU(3)$ breaking 
effects to $f_0(q_{\rm max}^2)$.

\begin{figure}[b]
\vspace{2mm}
\begin{center}

\includegraphics[width=0.5\linewidth,clip]{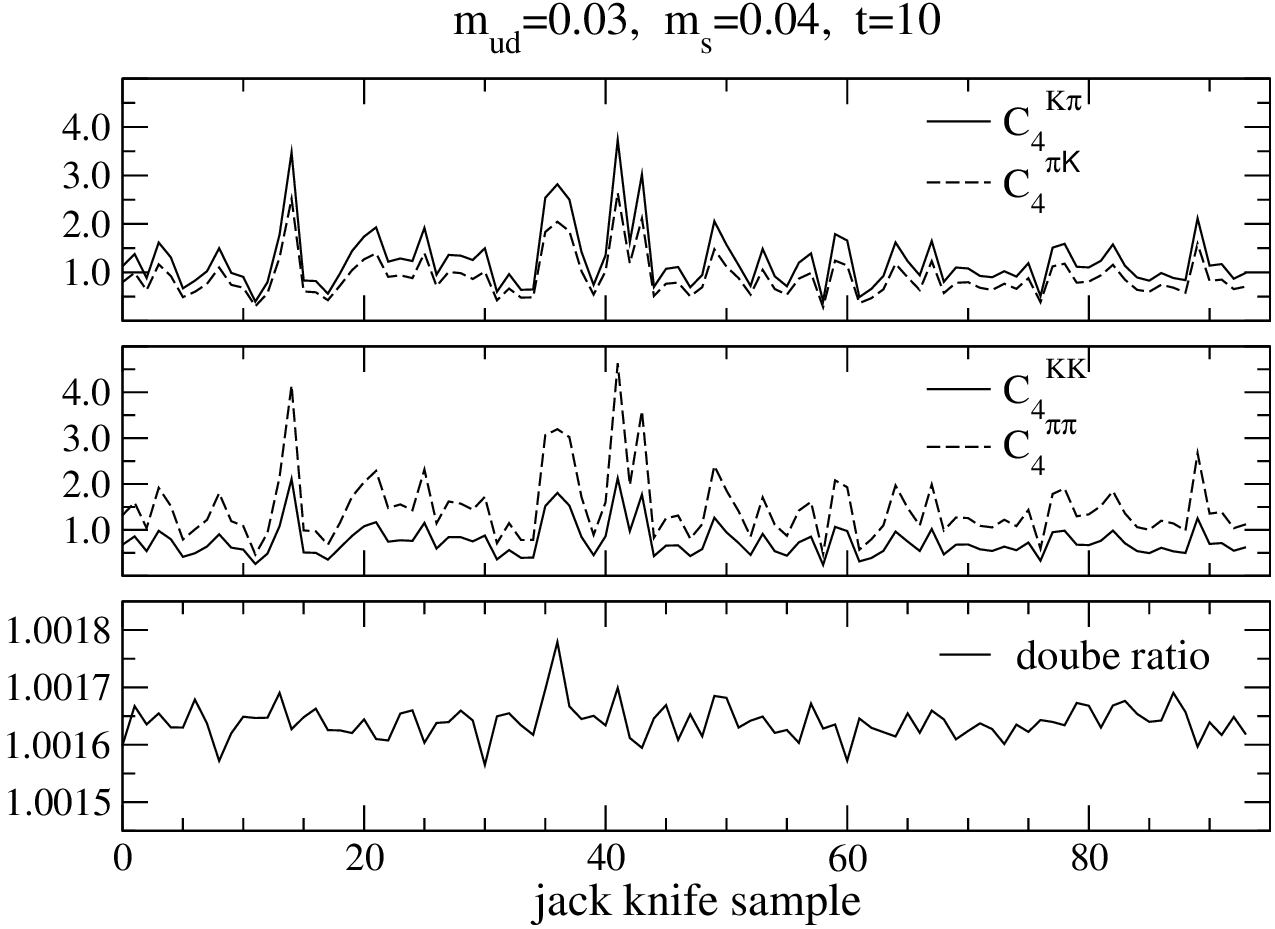}
\end{center}
\vspace{0mm}
\caption{
   Three-point functions (top and middle panels) and their double ratio $R$ 
   (bottom panel) for each jackknife sample 
   at $(m_{\rm ud},m_s)\!=\!(0.03,0.04)$.
}
\label{fig:ff:f0:dble_rat}
\end{figure}

In Fig.~\ref{fig:ff:f0:dble_rat}, we show the three-point functions
$C_4^{K\pi}$, $C_4^{\pi K}$, $C_4^{KK}$ and $C_4^{\pi \pi}$
for each jackknife sample.
Their fluctuation leads to the jackknife error of about 5\%.
As seen in the figure, they are highly correlated with each other, 
and hence the double ratio $R(t)$ has a very small fluctuation over the 
jackknife samples, which leads to the error of about 0.03\%.

\begin{figure}[b]
\vspace{2mm}
\begin{center}

\includegraphics[width=0.5\linewidth,clip]{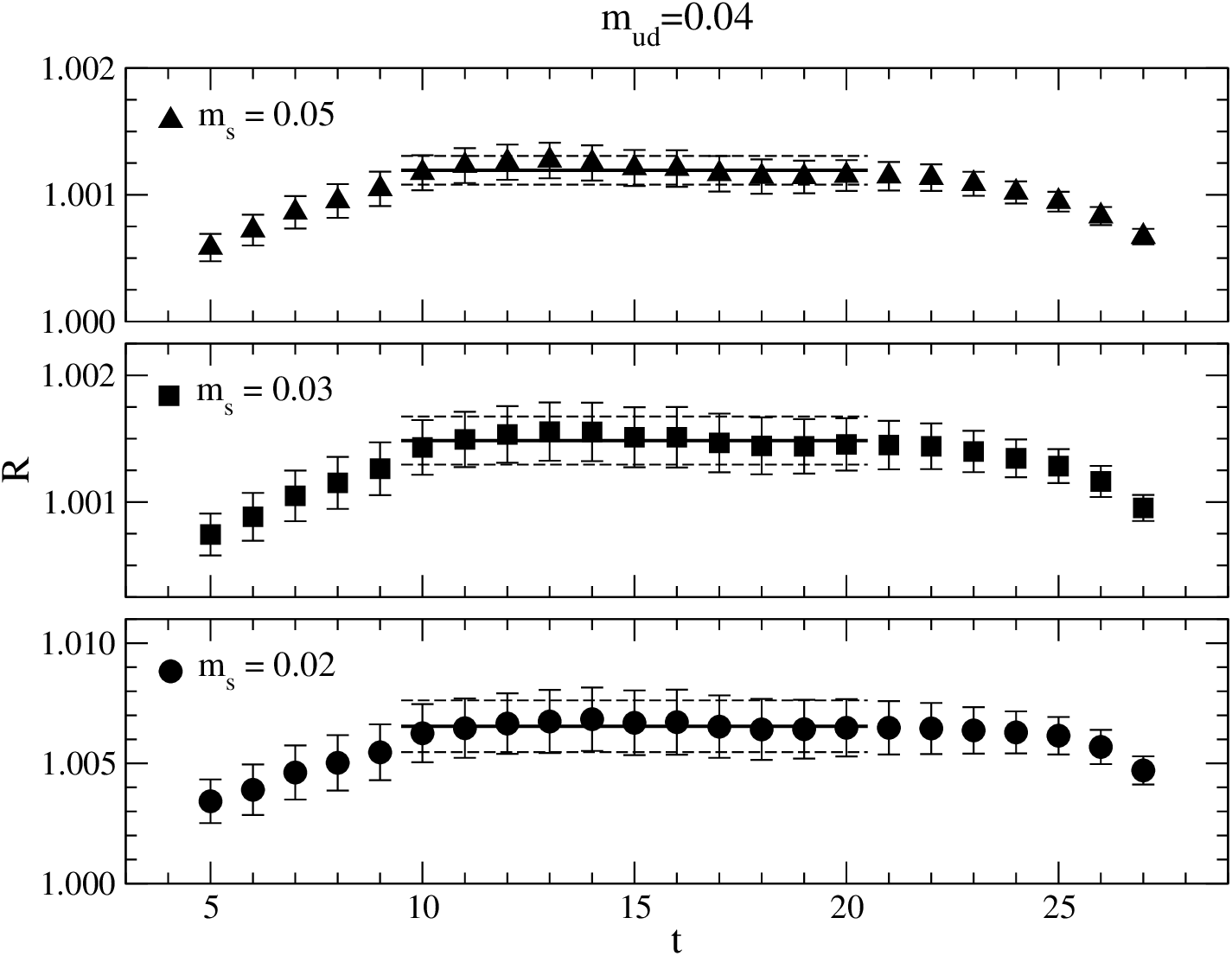}
\vspace{7mm}

\includegraphics[width=0.5\linewidth,clip]{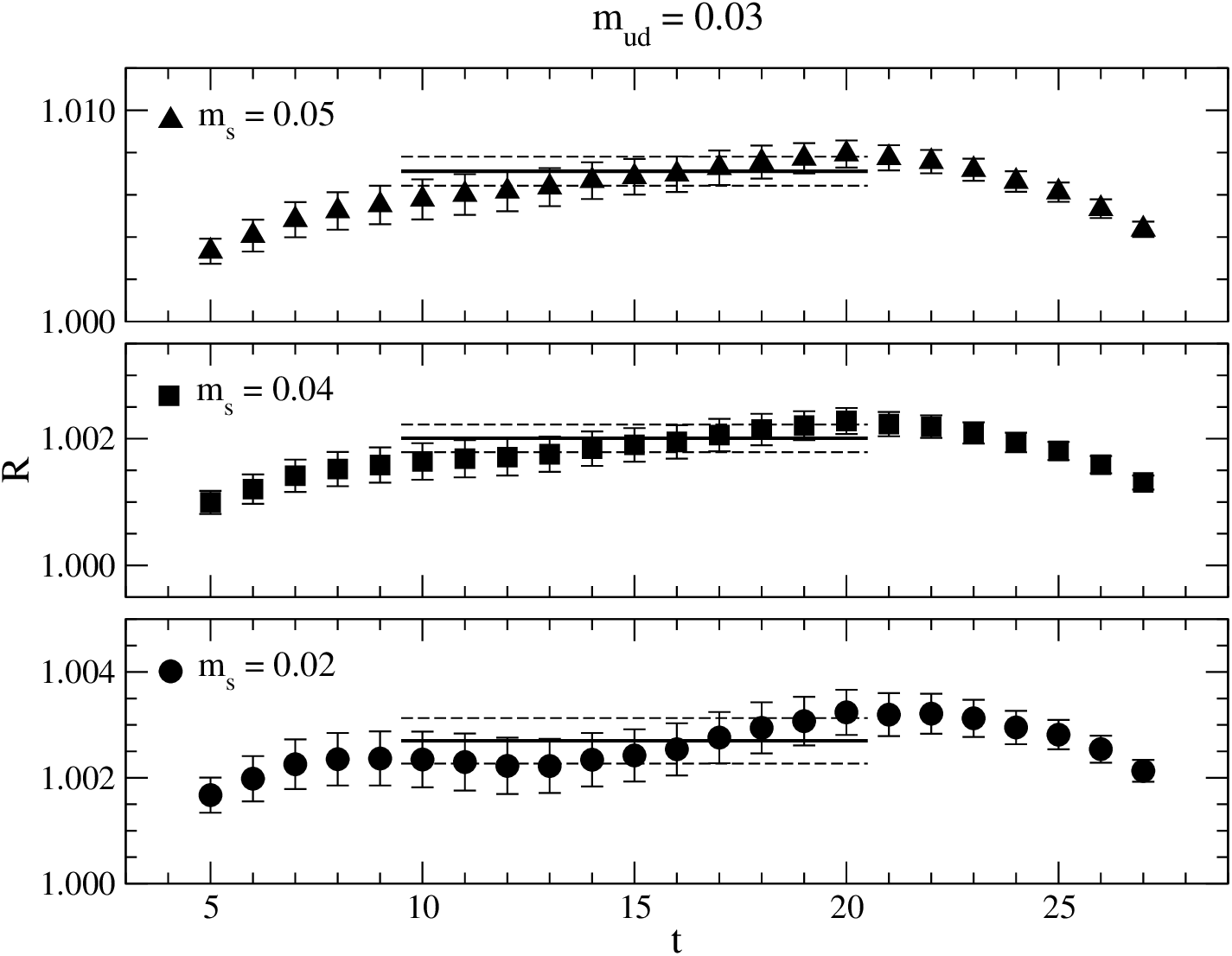}
\vspace{7mm}

\includegraphics[width=0.5\linewidth,clip]{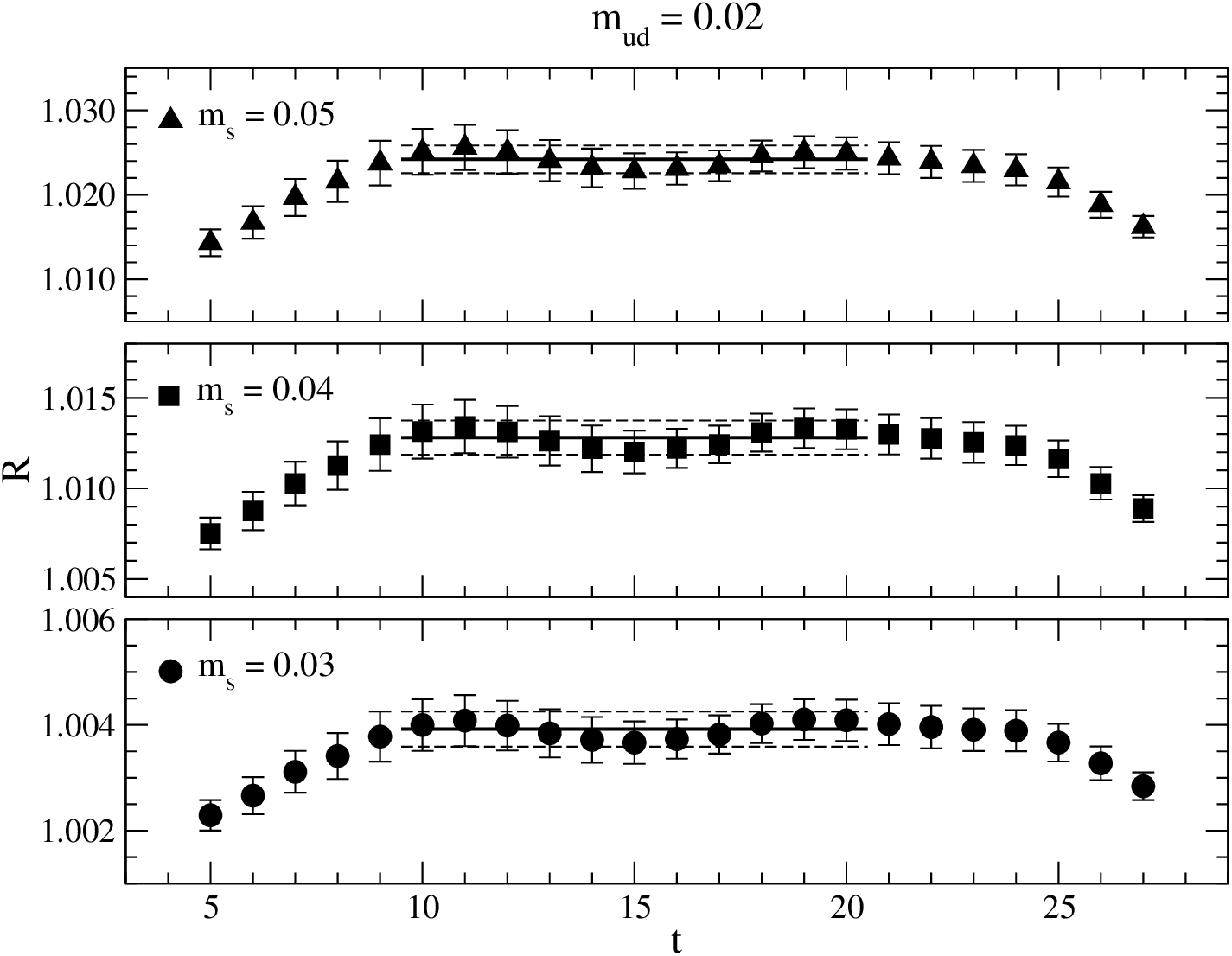}
\vspace{7mm}

\end{center}
\vspace{0mm}
\caption{
   Double ratio Eq.~(\ref{eqn:ff:f0:R14}) at $m_{ud}\!=\!0.04$ (top figure),
   0.03 (middle figure), and 0.02 (bottom figure).
}
\label{fig:ff:f0:R14}
\end{figure}

Figure~\ref{fig:ff:f0:R14} shows that the magnitude of the statistical error
does not change drastically at other values of $t$.
From the fitted value of $R(t)$ and measured meson masses on lattice,
we obtain $f_0(q_{\rm max}^2)$ summarized in Table~\ref{tbl:ff:f0:f0}.
There is a tendency that the error of $f_0(q_{\rm max}^2)$ increases
as $m_s$ deviates from $m_{ud}$, which is probably because $f_0(q_{\rm max}^2)$
deviates from its trivial value 1 towards larger $|m_s\!-\!m_{ud}|$. 
However, the statistical accuracy is $\lesssim$ 0.1\%
even in the worst case $(m_{ud},m_s)\!=\!(0.02,0.05)$.


\section{interpolation to zero momentum transfer}
\label{sec:q2_interp}


To study the $q^2$ dependence of the form factor, we calculate 
\bea
   F({\bf p},{\bf p}^{\prime})
   & = &
   \frac{f_+(q^2)}{f_0(q_{\rm max}^2)}
   \left( 1 + \frac{E_K({\bf p})-E_\pi({\bf p}^{\prime})}
                   {E_K({\bf p})+E_\pi({\bf p}^{\prime})} \, 
              \xi(q^2) 
   \right),
   \label{eqn:q2_interp:F24}
\eea
from a ratio 
\bea
   \tilde{R}(t;{\bf p},{\bf p}^{\prime})\,
   & = & 
   \frac{C_{4}^{K\pi}(t,t^{\prime};{\bf p},{\bf p}^{\prime})\,
         C^{K}(t;{\bf 0})\,C^{\pi}(t^{\prime}-t;{\bf 0})}
        {C_{4}^{K\pi}(t,t^{\prime};{\bf 0},{\bf 0})\,
         C^{K}(t;{\bf p})\,C^{\pi}(t^{\prime}-t;{\bf p}^{\prime})}
   \nn \\
   & \xrightarrow[t,(t^{\prime}-t) \to \infty]{} & 
   \frac{E_K({\bf p})+E_\pi({\bf p}^{\prime})}{M_K+M_\pi} \, 
   F({\bf p},{\bf p}^{\prime}).
   \label{eqn:q2_interp:R24}
\eea
Since 
$\tilde{R}(t;{\bf p},{\bf p}^{\prime})$ has its trivial value 1
at $|{\bf p}|\!=\!|{\bf p}^\prime|\!=\!0$, 
it might be a good probe to study how $f_0(q^2)$ changes 
as $q^2$ deviates from $q_{\rm max}^2$.
We also note that $\tilde{R}(t;{\bf p},{\bf p}^{\prime})$ 
is reduced to the double ratio employed in Ref.~\cite{kl3:fn:Nf2:JLQCD},
if we fix ${\bf p}\!=\!{\bf 0}$.

\begin{figure}[b]
\vspace{2mm}
\begin{center}
\includegraphics[width=0.5\linewidth,clip]{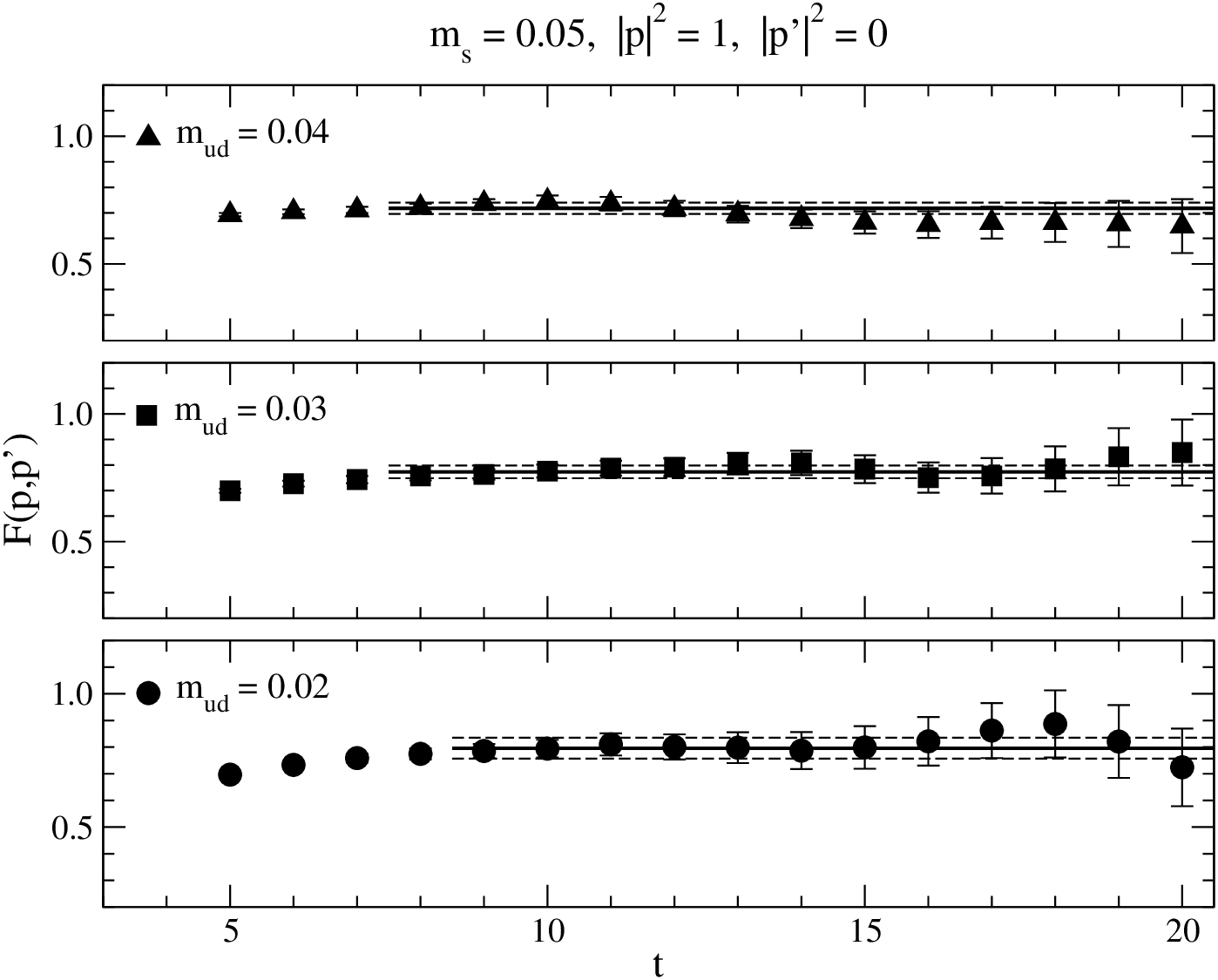}
\vspace{7mm}
  
\includegraphics[width=0.5\linewidth,clip]{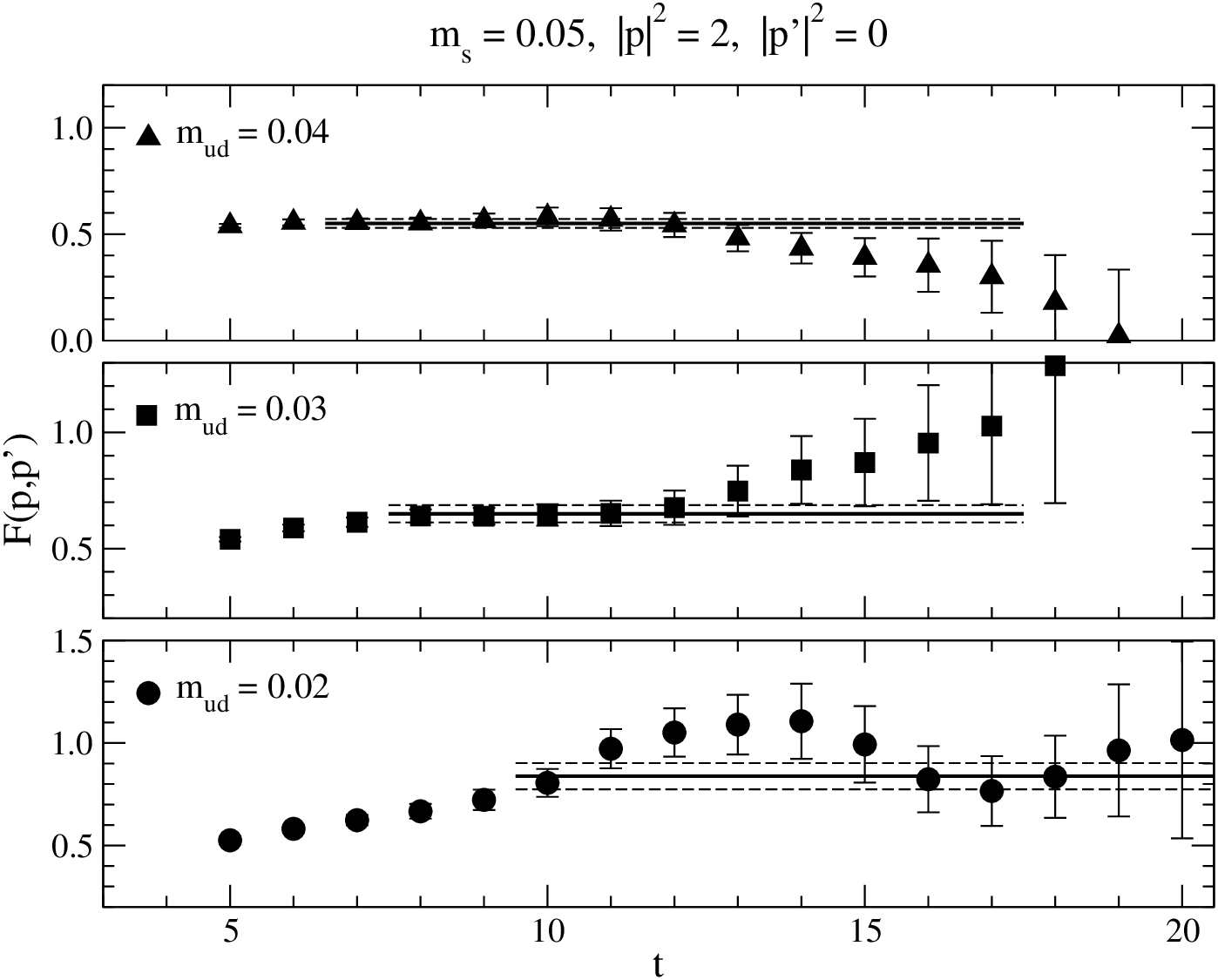}

\end{center}
\vspace{0mm}
\caption{
   Plots of $F({\bf p},{\bf p}^{\prime})$ defined 
   by Eq.(\ref{eqn:q2_interp:F24})
   with $m_s\!=\!0.05$ and $|{\bf p}^\prime|\!=\!0$.
   Top and bottom figures show data with $|{\bf p}|^2 \!=\! 1$ 
   and 2, respectively.
}
\label{fig:q2_interp:F24}
\end{figure}

In order to reduce the statistical error, 
we calculate $\tilde{R}(t;{\bf p},{\bf p}^{\prime})$ from 
three-point functions averaged over momentum configurations
which correspond to the same momentum sizes $|{\bf p}|$ and $|{\bf p}^\prime|$.
Figure~\ref{fig:q2_interp:F24} shows example of $F({\bf p},{\bf p}^{\prime})$ 
obtained from $\tilde{R}(t;{\bf p},{\bf p}^\prime)$ as a function of $t$.
We observe that data with the smallest non-zero momentum 
$|{\bf p}|^2\!=\!1$ show a clear plateau, and hence 
$F({\bf p},{\bf p}^{\prime})$ is determined with an accuracy of roughly 5\%.


In order to convert $F({\bf p},{\bf p}^{\prime})$ to $f_0(q^2)$, 
we evaluate $\xi(q^2)$ by employing the method 
proposed in Ref.\cite{kl3:fn:Nf0:italy}.
Namely, we measure the double ratio
\bea
    &&
    R_k(t;{\bf p},{\bf p}^{\prime})
    =
    \frac{C_k^{K\pi}(t,t^{\prime};{\bf p},{\bf p}^{\prime}) \,
          C_4^{KK}(t,t^{\prime};{\bf p},{\bf p}^{\prime})}
         {C_4^{K\pi}(t,t^{\prime};{\bf p},{\bf p}^{\prime}) \,
          C_k^{KK}(t,t^{\prime};{\bf p},{\bf p}^{\prime})},
    \hspace{5mm}
   (k=1,2,3),
   \label{eqn:q2_interp:R3k}
\eea
and calculate $\xi(q^2)$ from 
\bea
   &&
   \xi(q^2)
   = 
   \frac{-(E_K({\bf p})\!+\!E_K({\bf p}^{\prime}))\,
          (p\!+\!p^{\prime})_k
         +(E_K({\bf p})\!+\!E_\pi({\bf p}^{\prime}))\,
          (p\!+\!p^{\prime})_k \, R_k}
        { (E_K({\bf p})\!+\!E_K({\bf p}^{\prime}))\,
          (p\!-\!p^{\prime})_k
         -(E_K({\bf p})\!-\!E_\pi({\bf p}^{\prime}))\,
          (p\!+\!p^{\prime})_k \, R_k}.
	 \label{eqn:q2_interp:R3k_to_xi}
\eea
As for $\tilde{R}(t;{\bf p},{\bf p}^{\prime})$,
we first take the average for the relevant three-point functions 
over appropriately chosen momentum configuration and 
the Lorentz index for $V_{\mu}$ ($\mu\!=\!1,2,3$), 
and then double ratio $R_k(t;{\bf p},{\bf p}^{\prime})$ is constructed
from the averaged correlation function.
We note that $R_k(t;{\bf p},{\bf p}^{\prime})$ is exactly unity 
in the $SU(3)$ symmetric limit,
and is sensitive to $SU(3)$ breaking effects in 
the matrix element $C_\mu^{K\pi}$.

\begin{figure}[b]
\vspace{2mm}
\begin{center}
\includegraphics[width=0.5\linewidth,clip]{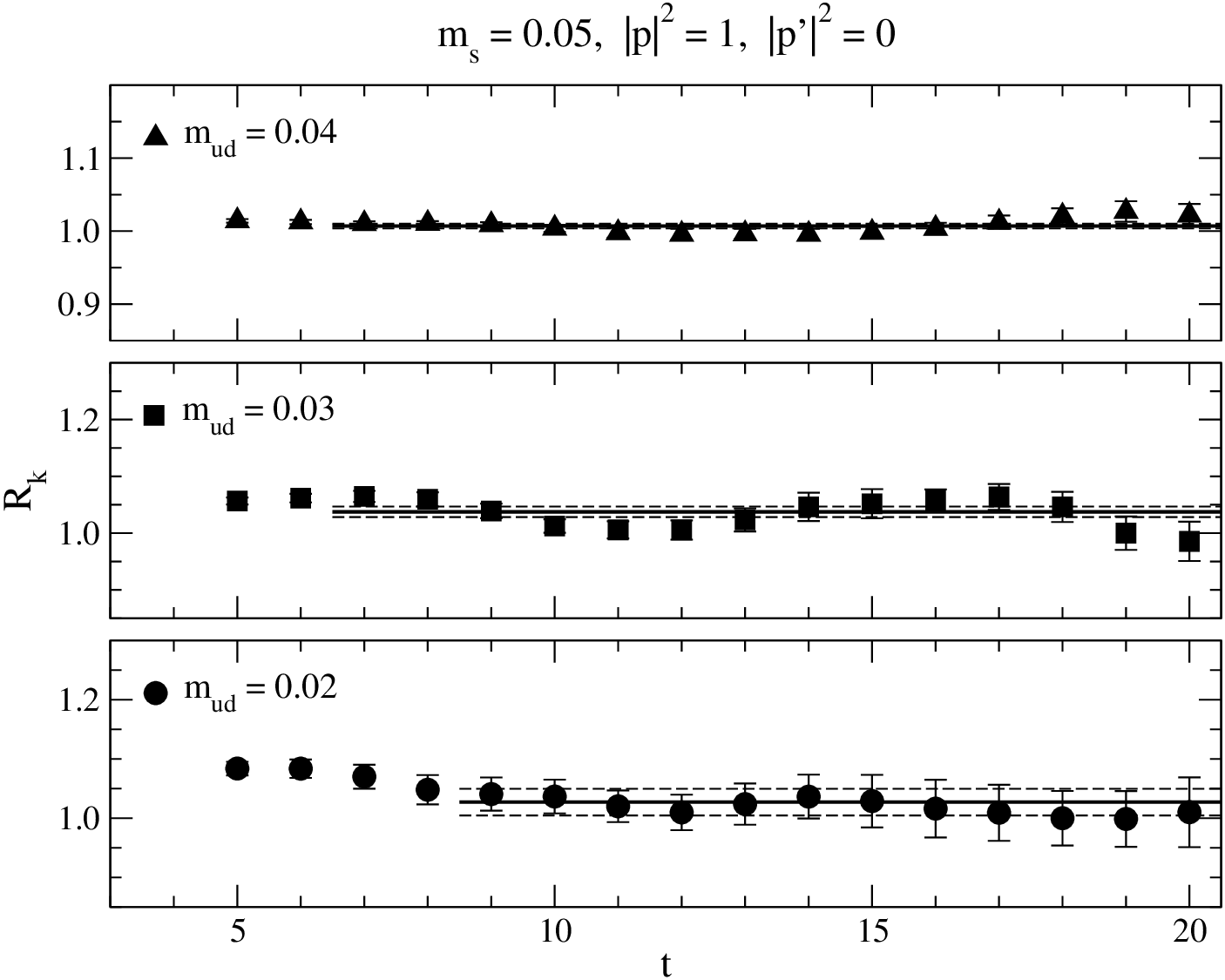}
\vspace{7mm}
  
\includegraphics[width=0.5\linewidth,clip]{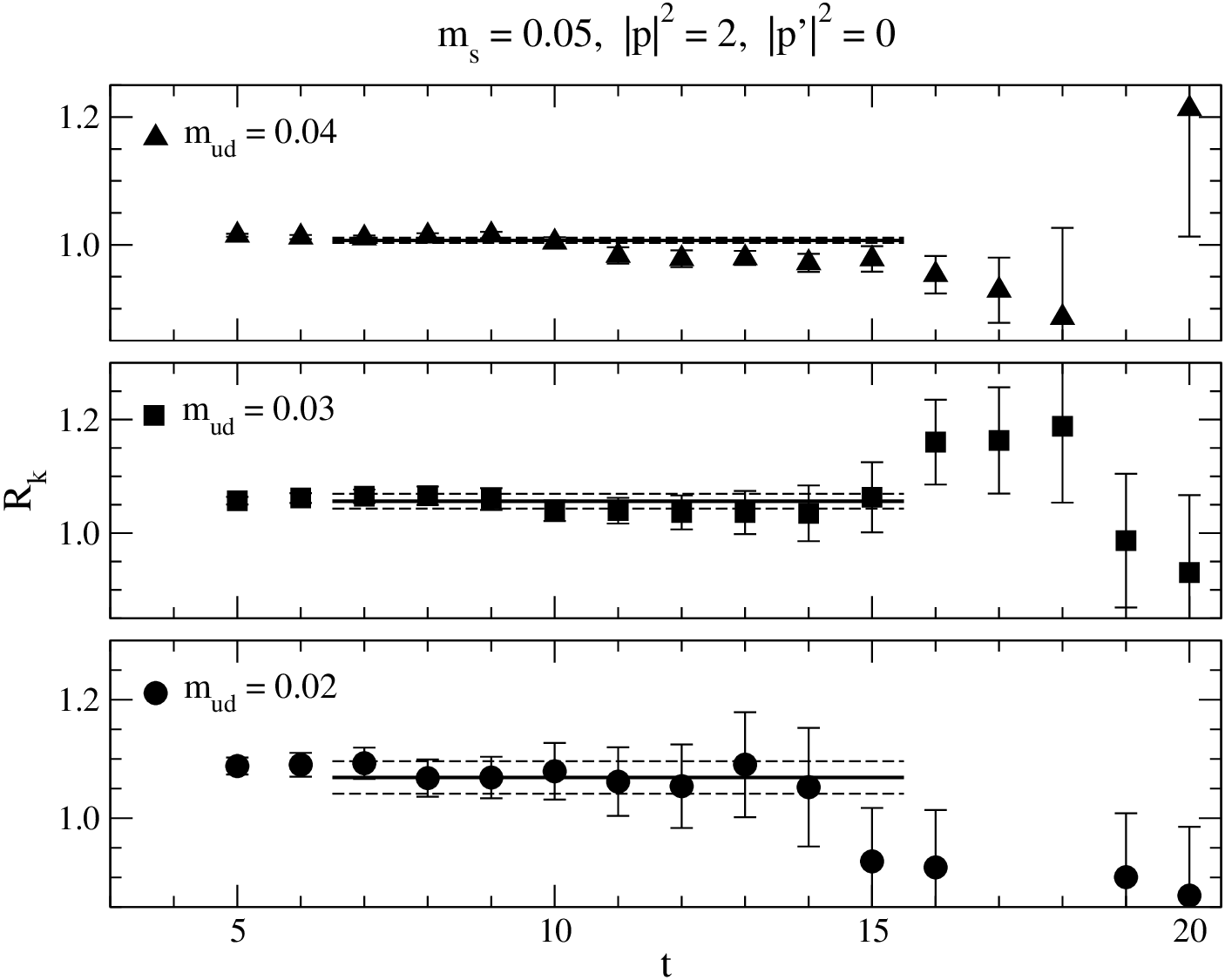}

\end{center}
\vspace{0mm}
\caption{
   Plots of double ratio $R_k(t;{\bf p},{\bf p}^{\prime})$ defined by 
   Eq.~(\ref{eqn:q2_interp:R3k})
   with $m_s\!=\!0.04$ and $|{\bf p}^\prime|\!=\!0$.
   Top and bottom figures show data with $|{\bf p}|^2 \!=\! 1$ 
   and 2, respectively.
}
\label{fig:q2_interp:F3k}
\end{figure}

Figure~\ref{fig:q2_interp:F3k} shows examples of 
$R_k(t;{\bf p},{\bf p}^{\prime})$ as a function of $t$.
We observe that, at most of our simulated quark masses,
$R_k(t;{\bf p},{\bf p}^{\prime})$ is close to unity, and hence
$\xi(q^2)$ from Eq.(\ref{eqn:q2_interp:R3k_to_xi})
has small magnitude $\lesssim \! 0.1$.
Its error is typically 30\%\,--\,100\% with our statistics.


We note that $F({\bf p},{\bf p}^{\prime})$ and $\xi(q^2)$ can be calculated 
also from the following ratios constructed from the $\pi \to K$ matrix element
\bea
   \tilde{R}^\prime(t;{\bf p},{\bf p}^{\prime})\,
   & = & 
   \frac{C_{4}^{\pi K}(t,t^{\prime};{\bf p},{\bf p}^{\prime})\,
         C^{\pi}(t;{\bf 0})\,C^{K}(t^{\prime}-t;{\bf 0})}
        {C_{4}^{\pi K}(t,t^{\prime};{\bf 0},{\bf 0})\,
         C^{\pi}(t;{\bf p})\,C^{K}(t^{\prime}-t;{\bf p}^{\prime})},
   \label{eqn:q2_interp:R24:pi2K}
   \\
   R_k^{\prime}(t;{\bf p},{\bf p}^{\prime})
   & = &
   \frac{C_k^{\pi K}(t,t^{\prime};{\bf p},{\bf p}^{\prime}) \,
         C_4^{KK}(t,t^{\prime};{\bf p},{\bf p}^{\prime})}
        {C_4^{\pi K}(t,t^{\prime};{\bf p},{\bf p}^{\prime}) \,
         C_k^{KK}(t,t^{\prime};{\bf p},{\bf p}^{\prime})}.
  \label{eqn:q2_interp:R3k:pi2K}
\eea
We confirm that, for $|{\bf p}|\!=\!0$ and $|{\bf p}^\prime|\!=\!1$, 
$\tilde{R}(t;{\bf p},{\bf p}^{\prime})$ and 
$\tilde{R}^\prime(t;{\bf p}^{\prime},{\bf p})$ 
give consistent results for $F({\bf p},{\bf p}^{\prime})$, 
while the latter leads to much smaller error. This is because, 
as described in Sec.\ref{sec:sim_param},
$C_{\mu}^{PQ}(t,t^{\prime};{\bf p},{\bf p}^{\prime})$ is measured with 
the single choice of the final meson momentum ${\bf p}^\prime$
for each $|{\bf p}^\prime|$,
and we can not take average of $\tilde{R}(t;{\bf p},{\bf p}^{\prime})$ 
over the momentum configuration ``\{${\bf p}^{\prime}$\}''.
We also observe that data with $|{\bf p}|,|{\bf p}^\prime|\!>\!0$
show poor signal.
Therefore, in the following analysis, we use 
$F({\bf p},{\bf p}^\prime)$ and $\xi(q^2)$ obtained 
from $\tilde{R}(t;{\bf p},{\bf p}^{\prime})$, 
$R_{k}(t;{\bf p},{\bf p}^{\prime})$,
$\tilde{R}^\prime(t;{\bf p},{\bf p}^{\prime})$, and 
$R_{k}^\prime(t;{\bf p},{\bf p}^{\prime})$ 
with $|{\bf p}^\prime|\!=\!0$.


In order to take the limit of zero momentum transfer of $f_0(q^2)$
reliably, we test two methods to calculate $f_0(0)$,
and check the consistency between results from these methods.
In the first method, which we call method-1 in the following,
we calculate $f_0(q^2)$ at simulated $q^2$ 
from $F({\bf p},{\bf p}^\prime)$ and $\xi(q^2)$. 
Then, the results for $f_0(q^2)$ and $f_0(q_{\rm max}^2)$ from 
$R(t)$ are interpolated to $q^2\!=\!0$. As well as the 
linear fitting form Eq.~(\ref{eqn:q2_interp:form:lin}), 
we test the quadratic form 
\bea
   f_0(q^2) & = & f_0(0)\,(1+\lambda_0^{(1)}\,q^2
                            +\lambda_0^{(2)}\,q^4),
   \label{eqn:q2_interp:form:quad} 
\eea
and the pole form
\bea
   f_0(q^2) & = & f_0(0)/(1-\lambda_0^{(1)}\,q^2).
   \label{eqn:q2_interp:form:pole} 
\eea
%

\begin{figure}[b]
\vspace{2mm}
\begin{center}
\includegraphics[width=0.5\linewidth,clip]{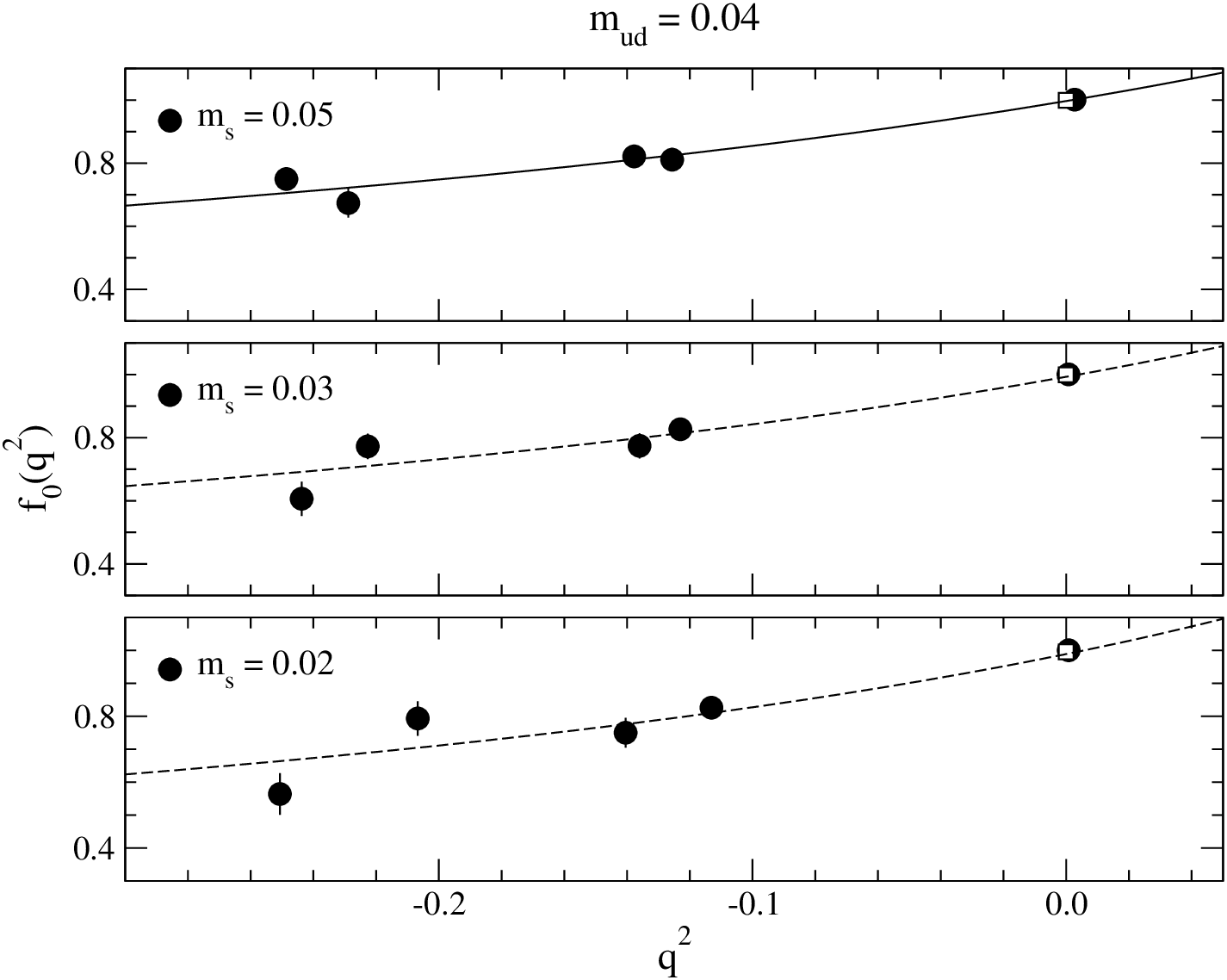}
\vspace{7mm}
  
\includegraphics[width=0.5\linewidth,clip]{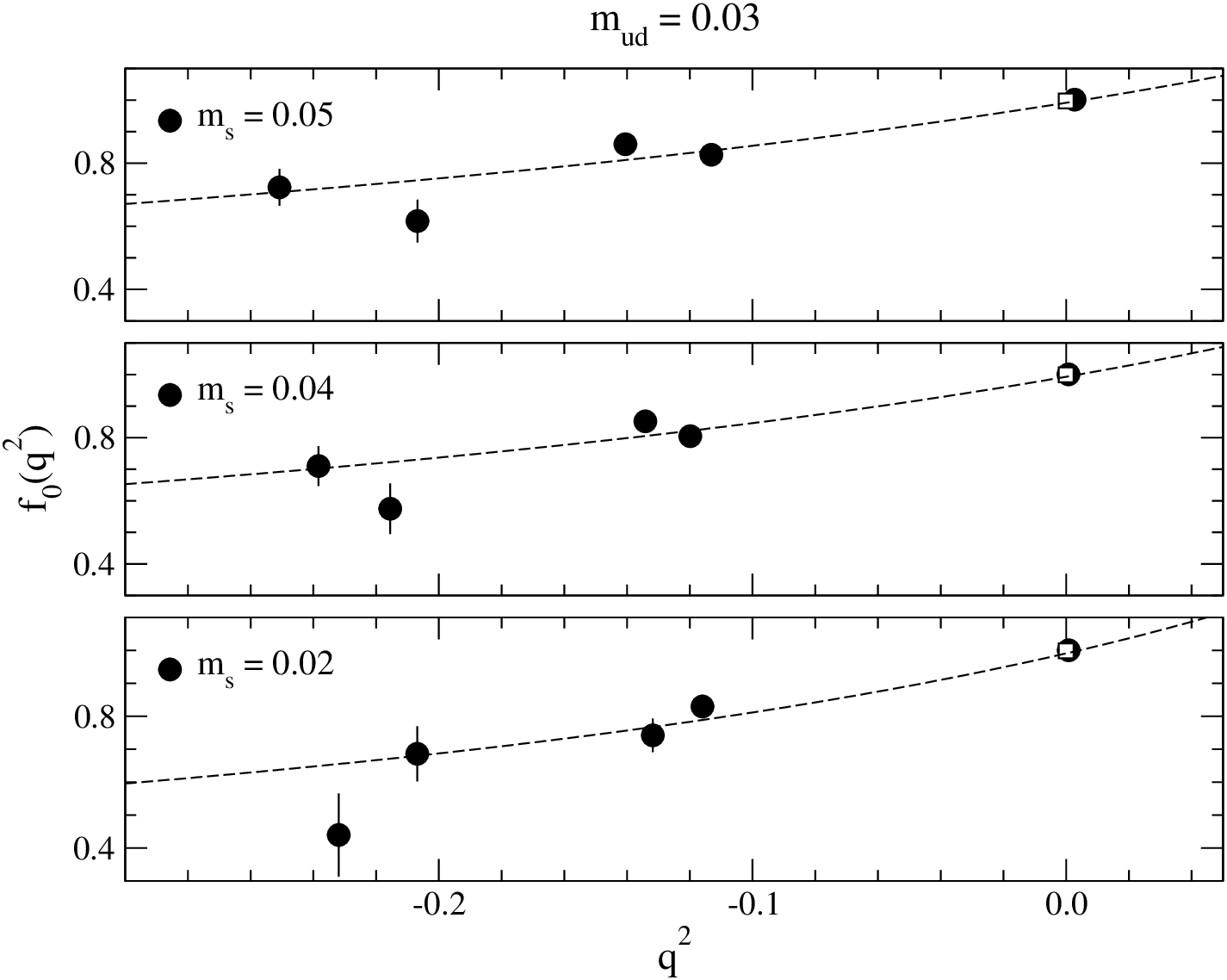}
\vspace{7mm}

\includegraphics[width=0.5\linewidth,clip]{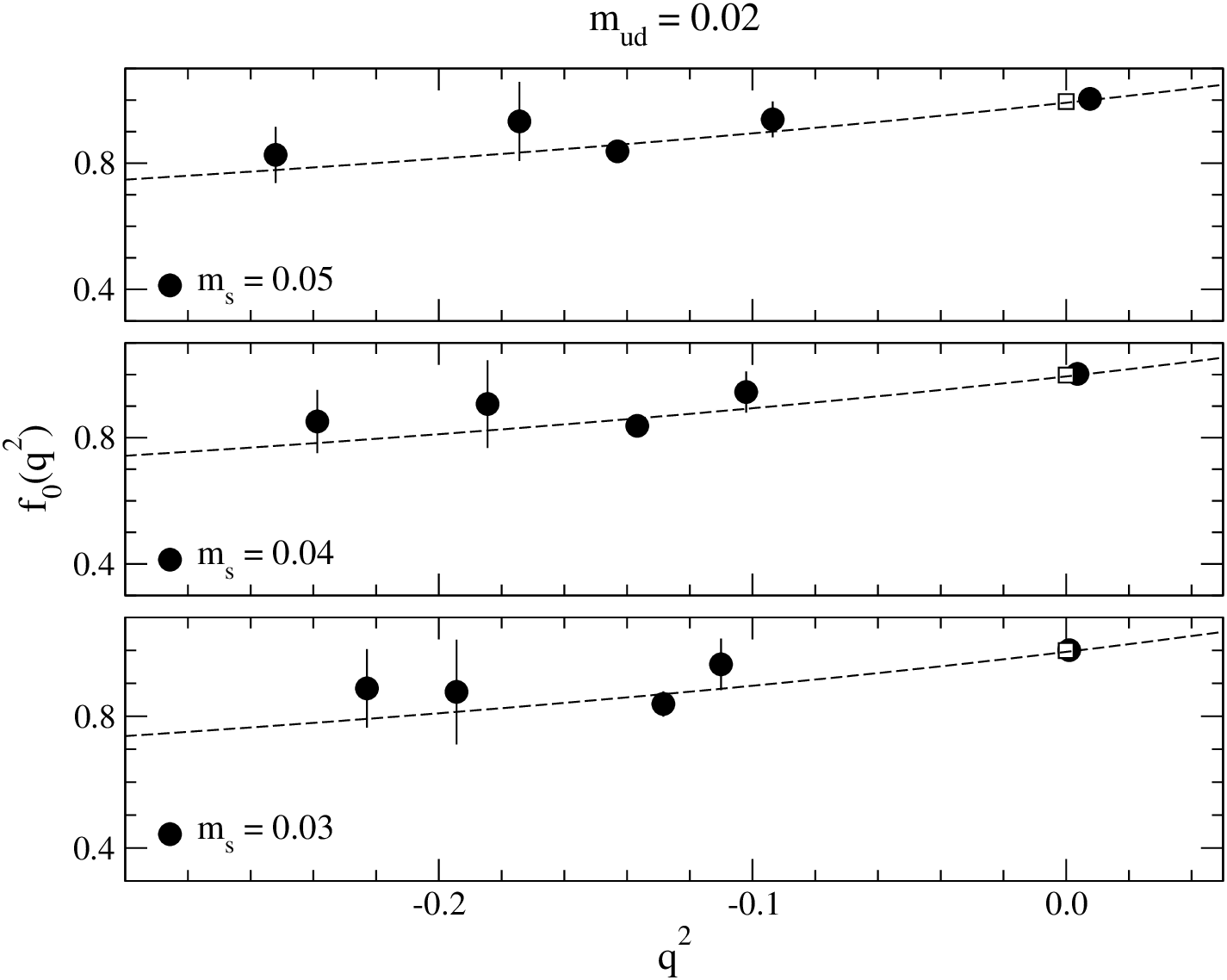}

\end{center}
\caption{
   Pole model interpolation of $f_0(q^2)$ to $q^2\!=\!0$.
   Top, middle and bottom figures show results at 
   sea quark mass $m_{ud}\!=\!0.04$, 0.03 and 0.02, respectively.
   Filled circles are $f_0(q^2)$ at simulated $q^2$, while 
   open squares represent interpolated value to $q^2\!=\!0$.
}
\label{fig:q2_interp:f0_vs_q2}
\end{figure}

The physical value of $q_{\rm max}^2$ for $K_{l3}$ decays is much smaller 
than that for $B$ meson decays. In addition, $q_{\rm max}^2$ is further 
reduced in our lattice calculation, since the simulated values of $m_{ud}$ 
are larger than its physical value.
As shown in Fig.~\ref{fig:q2_interp:f0_vs_q2}, 
$f_0(0)$ can be determined by a very short interpolation
from $q_{\rm max}^2$, where we have very accurate data 
$f_0(q_{\rm max}^2)$ from $R(t)$.
As a result, the choice of the interpolation form does not affect 
the interpolated value $f_0(0)$ significantly. 
Actually we observe that the interpolated values obtained from 
the three forms, 
Eqs.~(\ref{eqn:q2_interp:form:lin}), (\ref{eqn:q2_interp:form:quad})
and (\ref{eqn:q2_interp:form:pole})
are consistent with each other.


In the following, we employ the result from the pole form
Eq~(\ref{eqn:q2_interp:form:pole}), because i) 
data of $f_0(q^2)$ does not have a strong curvature,
and hence the quadratic fit leads to ill-determined $\lambda_0^{(2)}$,
which has typically 100\% error,
and ii) the pole form leads to slightly 
smaller value of $\chi^2/\mbox{dof}$ than the linear fit.
The fit parameters are summarized in Table~\ref{tbl:q2_interp:mthd1}.
While the pole form at $m_{ud}\!=\!0.03$
leads to a slightly higher value of $\chi^2/\mbox{dof}$
than at other values of $m_{ud}$,
the inclusion of the quadratic term does not reduce $\chi^2/\mbox{dof}$
significantly.
We note that the statistical error on $f_0(0)$ is $\lesssim 0.3$\% level.


We also test an alternative method to calculate $f_0(0)$ employed in 
Ref.\cite{kl3:fn:Nf2:JLQCD}, which we call method-2 in the following.
%
%
In this method,
we first take the limit of $F({\bf p},{\bf p}^{\prime})$ and $\xi(q^2)$ 
to $q^2\!=\!0$, 
and then calculate $f_0(0)$ from $\left. 
F({\bf p},{\bf p}^{\prime})\right|_{q^2=0}$ and $\xi(0)$.
Since $F({\bf p},{\bf p}^{\prime})$ depends on two momenta $|{\bf p}|$,
and $|{\bf p}^\prime|$, 
the $q^2$ interpolation of $F({\bf p},{\bf p}^{\prime})$ has to be carried out 
using data with fixed $|{\bf p}|$ (or $|{\bf p}^{\prime}|$).
This also enables us to identify $|{\bf p}^{\prime}|$ $(|{\bf p}|)$ 
corresponding to $q^2\!=\!0$, which is needed to convert 
$F({\bf p},{\bf p}^{\prime})|_{q^2\!=\!0}$ to $f_0(0)$.
In the following, we repeat the interpolation for two data sets 
with $|{\bf p}|\!=\!0$ and $|{\bf p}^{\prime}|\!=\!0$, 
and take the average of results for $\left. 
F({\bf p},{\bf p}^{\prime})\right|_{q^2=0}$.

\begin{figure}[b]
\vspace{2mm}
\begin{center}

\includegraphics[width=0.5\linewidth,clip]{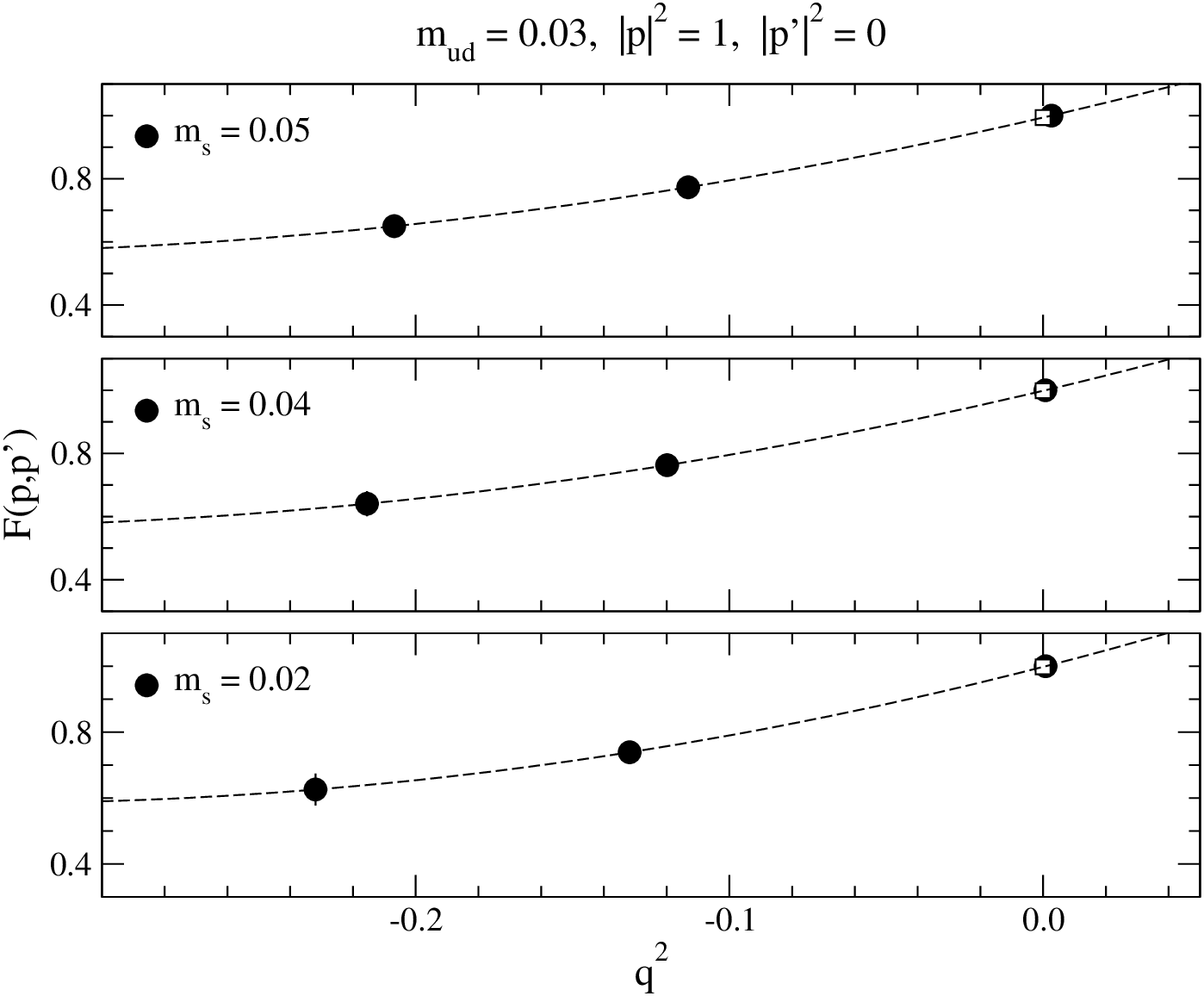}
\vspace{5mm}

\includegraphics[width=0.5\linewidth,clip]{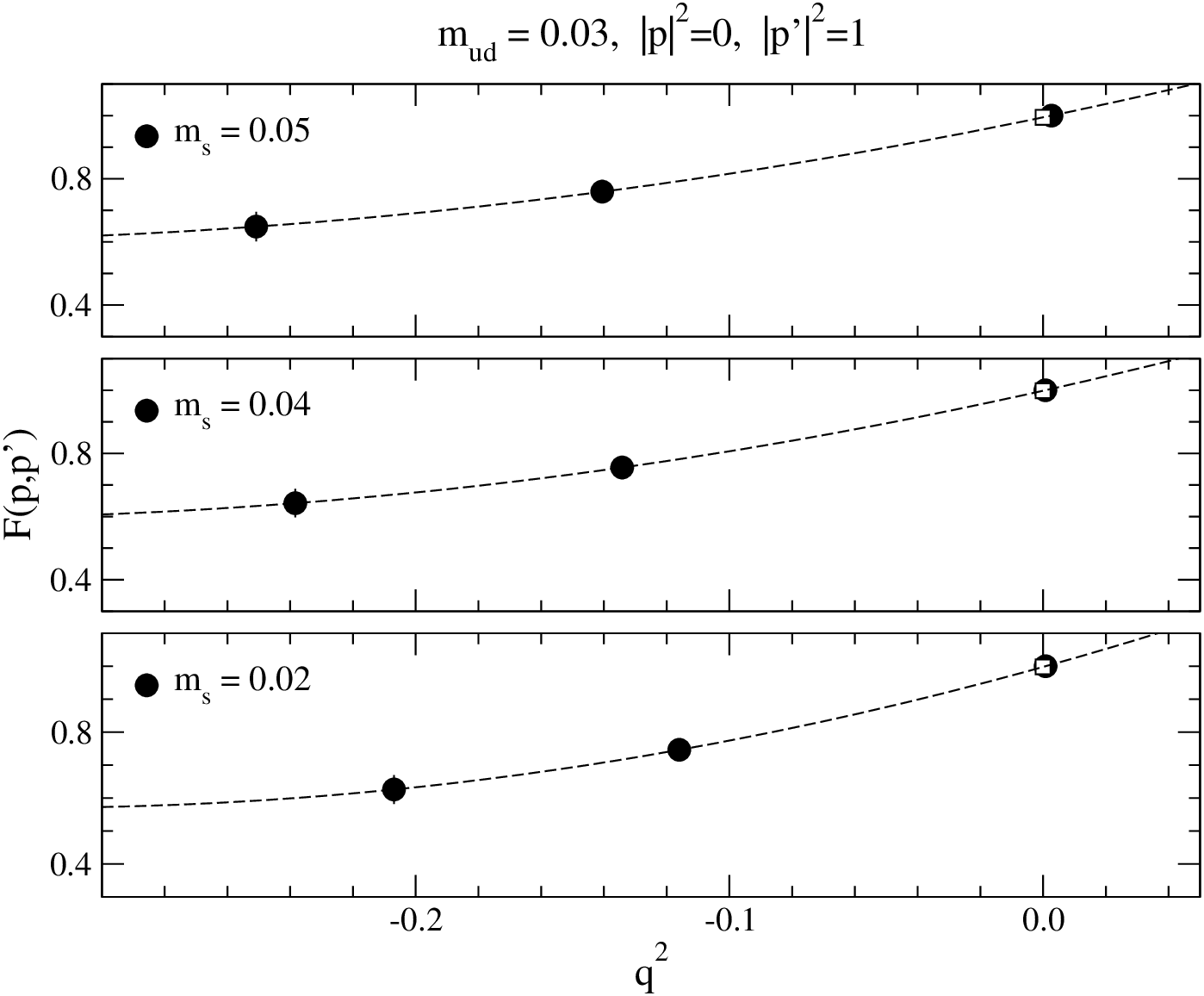}

\end{center}
\vspace{0mm}
\caption{
   Interpolation of $F({\bf p},{\bf p}^{\prime})$ as a function of $q^2$
   at $m_{ud}\!=\!0.03$.
   Top and bottom figures show data with $|{\bf p}^\prime|\!=\!0$ 
   and $|{\bf p}|\!=\!0$, respectively.
}
\label{fig:q2_interp:F24_vs_q2}
\end{figure}

For the $q^2$ interpolation, 
we test linear, pole, and quadratic fitting forms similar to 
Eqs.~(\ref{eqn:q2_interp:form:lin}), (\ref{eqn:q2_interp:form:quad})
and (\ref{eqn:q2_interp:form:pole}),
and employ the quadratic fit
\bea
   \left. F({\bf p},{\bf p}^{\prime}) \right|_{q^2}
   & = &
   \left. F({\bf p},{\bf p}^{\prime}) \right|_{q^2=0}
   \cdot (1+c_{F,1}\, q^2 + c_{F,2}\,q^4)
   \hspace{5mm}
   (\mbox{$|{\bf p}|$ or $|{\bf p}|^\prime$ is fixed}),
   \label{eqn:q2_interp:F24_vs_q2}
\eea
since this fit leads to the smallest value for $\chi^2/\mbox{dof}$
among the tested forms, and to the reasonably well-determined $c_{F,2}$.
Examples of this quadratic fit are shown in 
Fig.~\ref{fig:q2_interp:F24_vs_q2}.

\begin{figure}[b]
\vspace{2mm}
\begin{center}

\includegraphics[width=0.5\linewidth,clip]{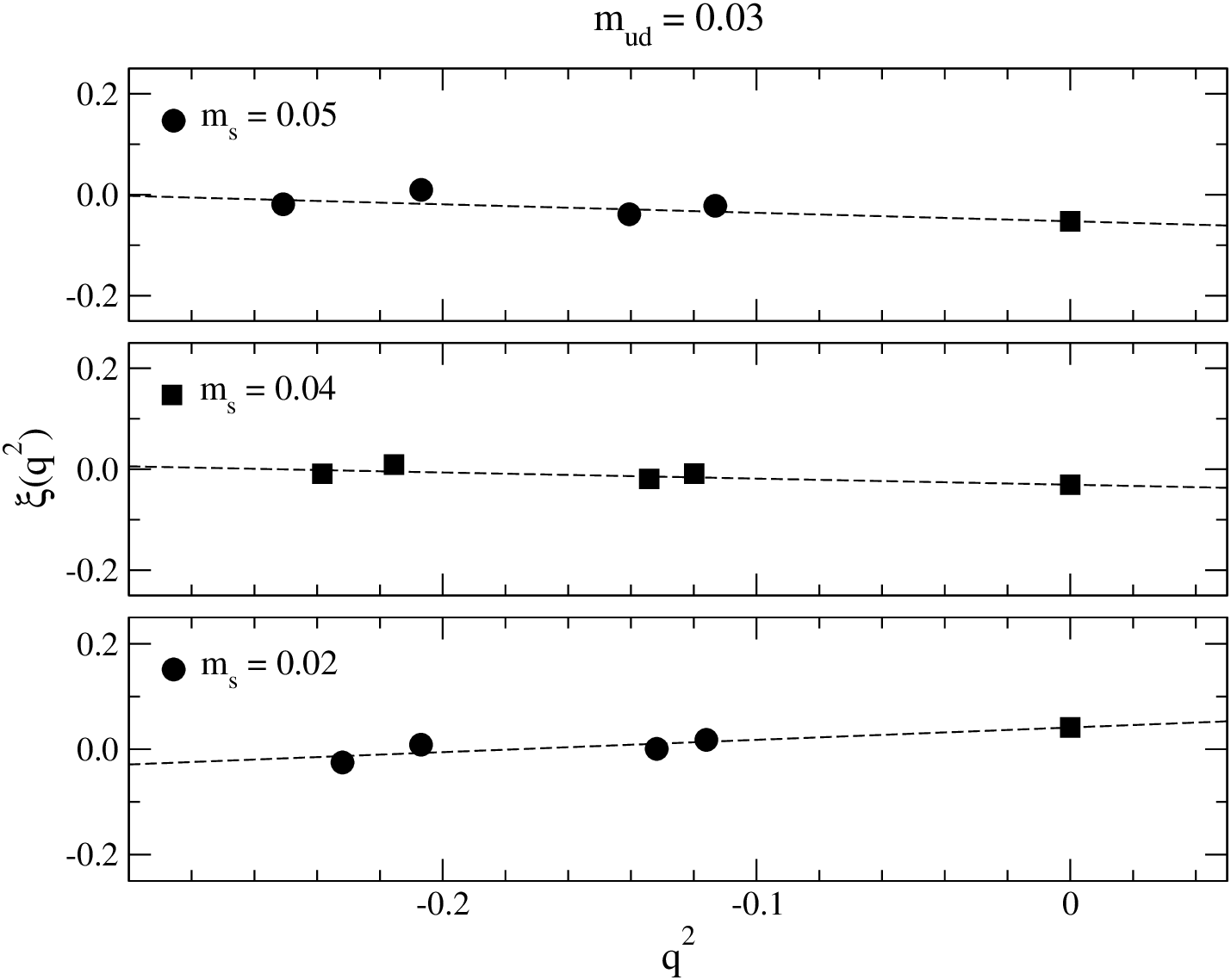}

\end{center}
\vspace{0mm}
\caption{
   Linear extrapolation of $\xi(q^2)$ as a function of $q^2$
   at $m_{ud}\!=\!0.03$.
}
\label{fig:q2_interp:xi_vs_q2}
\end{figure}

While $\xi(0)$ has to be determined by an extrapolation, 
we observe that $\xi(q^2)$ has very mild $q^2$ dependence,
as seen in Fig.~\ref{fig:q2_interp:xi_vs_q2},
and the simplest linear form
\bea
   \xi(q^2) 
   & = &
   \xi(0) \, (1+c_{\xi,1}\, q^2)
   \label{eqn:q2_interp:xi_vs_q2}
\eea
leads to reasonably small $\chi^2/\mbox{dof}$.
Table~\ref{tbl:q2_interp:mthd2} shows $\xi(0)$ and 
$f_0(0)$ calculated from $\left. F({\bf p},{\bf p}^{\prime})\right|_{q^2=0}$
and $\xi(0)$.


We observe that two methods of the $q^2$ interpolation give consistent results 
for $f_0(0)$ with each other. However, the error from method-2 is slightly 
larger than from method-1
due to the error for $\xi(0)$ enhanced by the extrapolation.
Therefore, we employ results from method-1 in the following.

We note that the above observation is opposite from that in our preliminary
study \cite{kl3:fn:Nf2:RBC}, where method-1 give larger error for 
$f_0(0)$ mainly due to the uncertainty in $\xi(q^2)$ at simulated $q^2$.
As described in Sec.\ref{sec:sim_param}, we calculate correlation functions
with the different choice of the smearing function 
from that in Ref.\cite{kl3:fn:Nf2:RBC}.
This improves accuracy in $\xi(q^2)$ at simulated $q^2$ and hence
$f_0(0)$ in method-1.
However, in order to reduce the uncertainty from method-2, the change 
of the smearing functions is not sufficient and we need to have data 
at small $q^2$ for a better control of the extrapolation of $\xi(q^2)$.


\section{Chiral extrapolation}
\label{sec:chiral_fit}

In the chiral extrapolation of $f_+(0)\!=\!f_0(0)$,
we rewrite the ChPT expansion Eq.~(\ref{eqn:kl3:f+:expand}) as
\bea
   f_+(0) = 1 + f_2 + \Delta f,
   \label{eqn:chiral_fit:f+_expand}
\eea 
where $\Delta f$ represents all higher order corrections
starting at $O(M_{K,\pi,\eta}^4)$.


As mentioned in Sec.\ref{sec:kl3}, 
the leading correction $f_2$ does not have analytic terms 
from the $O(p^4)$ chiral Lagrangian thanks to the Ademollo-Gatto theorem.
It is shown in Ref.\cite{kl3:f2:pqQCD} that the above statement
is true even in two-flavor partially quenched (PQ) theory, as seen in 
their PQChPT formula of $f_2$ 
\bea
   f_2^{\rm (PQ)}
   & = &
   -\frac{2\,M_K^2+M_\pi^2}{32\,\pi^2\,f_\pi^2}
   -\frac{3\,M_K^2\,M_\pi^2 \mbox{ln}[M_\pi^2/M_K^2]}
         {64\,\pi^2\,f_\pi^2\,(M_K^2-M_\pi^2)}
   \nn \\
   &   &
   +\frac{M_K^2\,(4\, M_K^2-M_\pi^2)\, \mbox{ln}[2-M_\pi^2/M_K^2]}
         {64\,\pi^2\,f_\pi^2\,(M_K^2-M_\pi^2)}.
   \label{eqn:chiral_fit:f2:pqchpt}
\eea
Therefore, $f_2$ does not contain any poorly-known LECs
in the $O(p^4)$ chiral Lagrangian, and its value at simulated
quark mass can be precisely calculated from the measured meson masses
through Eq.~(\ref{eqn:chiral_fit:f2:pqchpt}).


Consequently, the chiral extrapolation of $f_+(0)$ is nothing but 
the extrapolation of the higher order correction $\Delta f$.
Since it is also proportional to $(m_s-m_{ud})^2$ thanks to the 
Ademollo-Gatto theorem, we consider the ratio
\bea
   R_{\Delta f}
   & = & 
   \frac{\Delta f}{(M_K^2-M_\pi^2)^2},
\eea     
as in Ref.\cite{kl3:fn:Nf0:italy}, and extrapolate it to the physical 
quark mass. To this end, we test the following constant, 
linear and quadratic fits
\bea
   R_{\Delta f} 
   & = & 
   c_0,
   \label{eqn:chiral_fit:fit_form:const}
   \\
   R_{\Delta f} 
   & = & 
   c_0 + c_{1,v}\,(M_K^2+M_\pi^2),
   \label{eqn:chiral_fit:fit_form:val-lin}
   \\
   R_{\Delta f} 
   & = & 
   c_0 + c_{1,s}\,M_\pi^2 + c_{1,v}\,(M_K^2+M_\pi^2),
   \label{eqn:chiral_fit:fit_form:lin}
   \\
   R_{\Delta f} 
   & = & 
   c_0 + c_{1,s}\,M_\pi^2 + c_{1,v}\,(M_K^2+M_\pi^2) 
       + c_{2,s}\,M_\pi^4 + c_{2,v}\,(M_K^2+M_\pi^2)^2.
   \label{eqn:chiral_fit:fit_form:quad}
\eea
The constant fit Eq.~(\ref{eqn:chiral_fit:fit_form:const}) 
should work if $\Delta f$ is dominated by the analytic term in $f_4$.
Linear and quadratic dependences in Eqs.~(\ref{eqn:chiral_fit:fit_form:val-lin})\,--\,(\ref{eqn:chiral_fit:fit_form:quad}) are assumptions
for an effective description of the chiral logarithms 
in $f_4$ and higher order corrections.

\begin{figure}[b]
\vspace{2mm}
\begin{center}

\includegraphics[width=0.5\linewidth,clip]{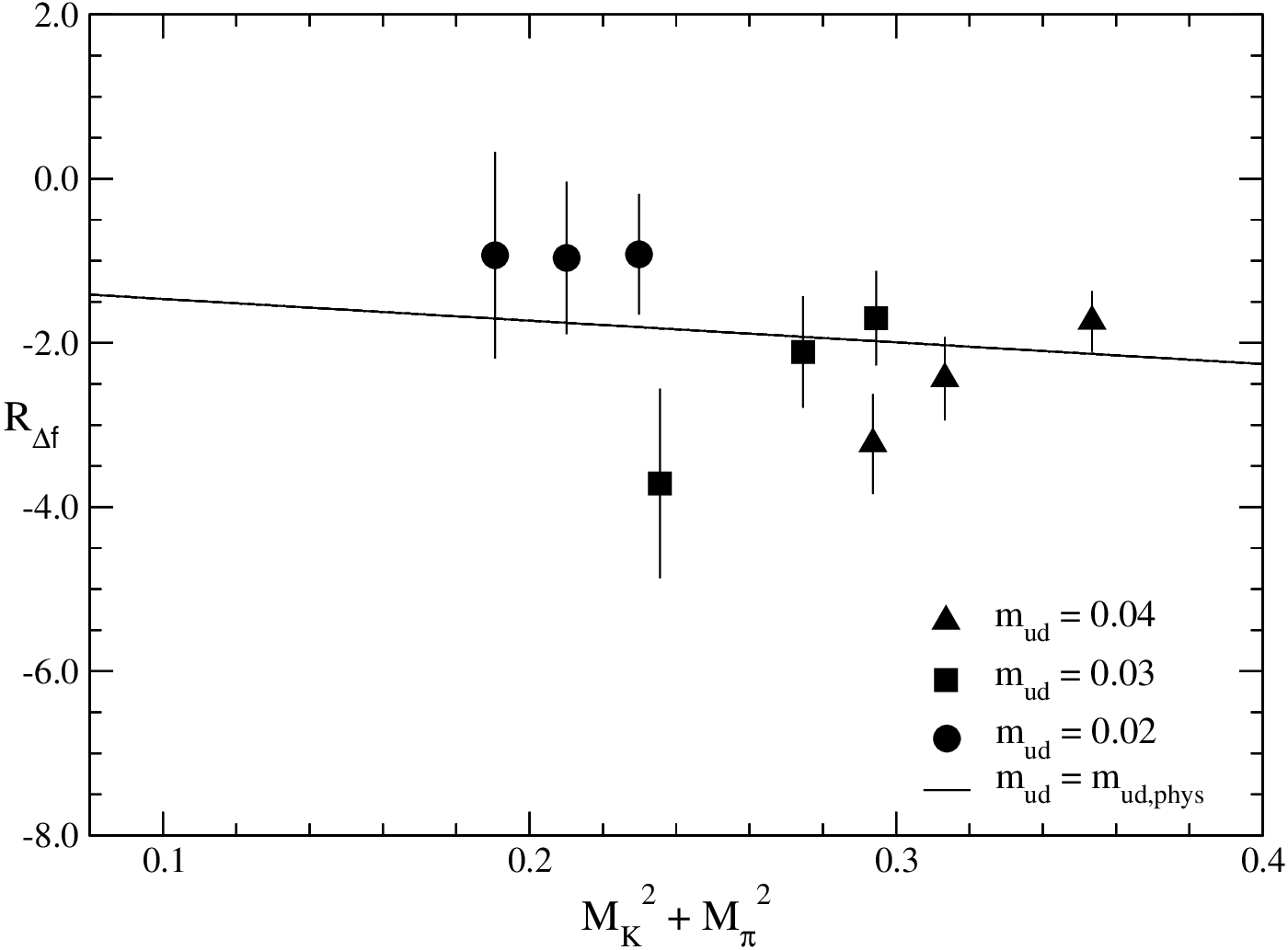}
\vspace{5mm}

\includegraphics[width=0.5\linewidth,clip]{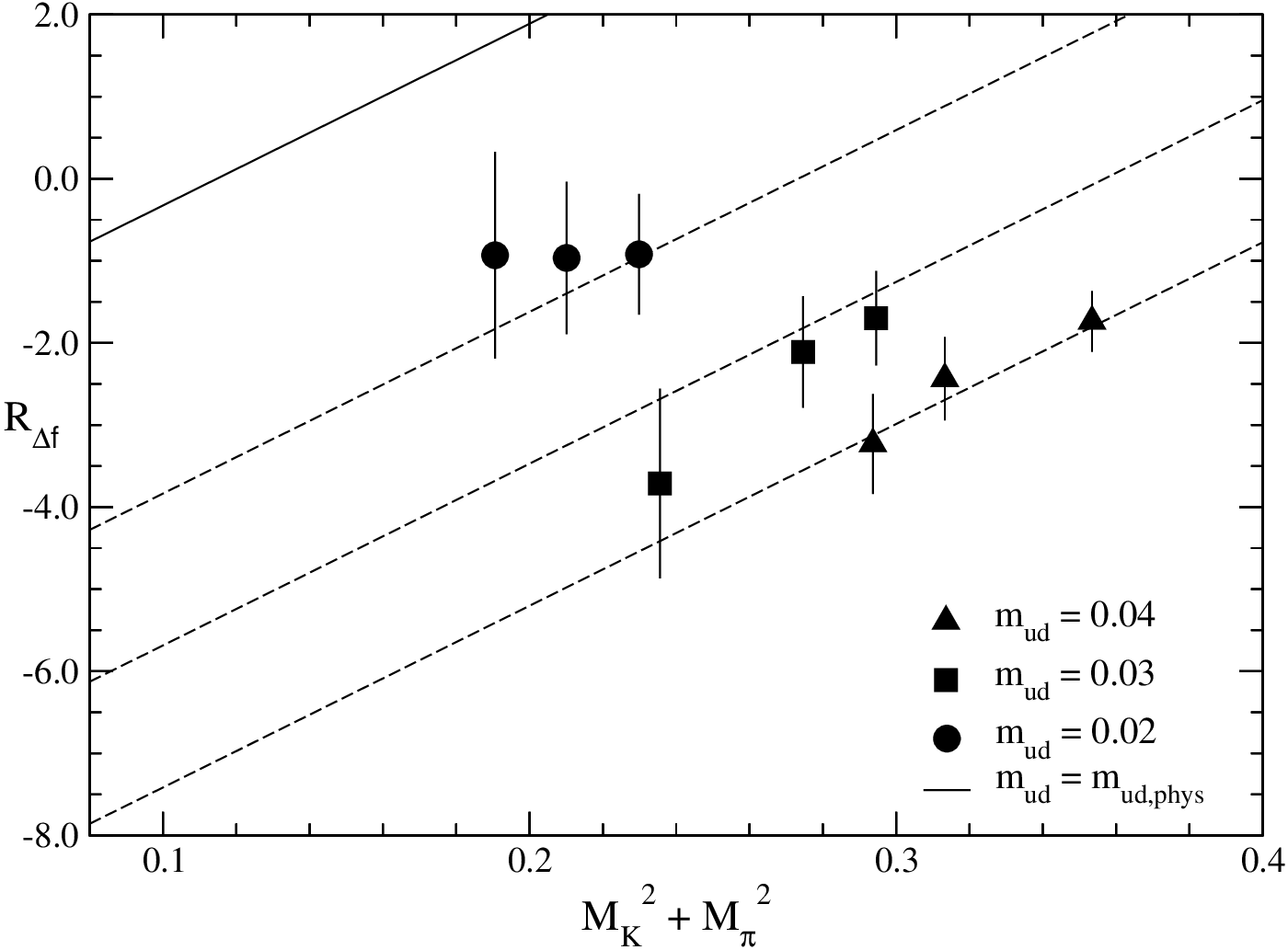}

\end{center}
\vspace{0mm}
\caption{
   Chiral extrapolation of $R_{\Delta f}$.
   Top and bottom figures show results using 
   Eqs.(\ref{eqn:chiral_fit:fit_form:val-lin}) and 
   (\ref{eqn:chiral_fit:fit_form:lin}), 
   respectively.
}
\label{fig:chiral_fit:f+}
\end{figure}

Figure~\ref{fig:chiral_fit:f+} shows the chiral extrapolation
using the linear forms Eqs.~(\ref{eqn:chiral_fit:fit_form:val-lin}) and 
(\ref{eqn:chiral_fit:fit_form:lin}). 
We observe that $R_{\Delta f}$ has mild dependence on 
the sea and valence quark masses, and the linear and even 
constant fits achieve sufficiently small value of $\chi^2/\mbox{dof}$.
While the quadratic fit Eq.~(\ref{eqn:chiral_fit:fit_form:quad}) also 
gives a small value of $\chi/\mbox{dof}$,
it leads to more than 100\% error for both of $c_{2,s}$ and $c_{2,v}$.
We, therefore, do not use the results from the quadratic fit 
in the following discussion.


From the fit parameters summarized in Table~\ref{tbl:chiral_fit:f+}
and the physical meson masses determined 
in Ref.\cite{Spectrum+BK:Nf2QCD:RBC},
we obtain $\Delta f$ at the physical quark mass which is also 
collected in Table~\ref{tbl:chiral_fit:f+}.
We note that all fits lead to consistent results for $\Delta f$ 
with each other.
We obtain 
\bea
   \Delta f = -0.009(9)(6),
   \label{eqn:chiral_fit:delta_f}
\eea
by employing result from the linear fit 
Eq.~(\ref{eqn:chiral_fit:fit_form:lin}), which is also employed 
in the unquenched calculations in Refs.\cite{kl3:fn:Nf2:JLQCD,kl3:fn:Nf3:FNAL}.
The first error is statistical, and the second is a systematic
error due to the chiral extrapolation which is estimated as 
the largest deviation in $\Delta f$ among the constant and linear fits.


By using $f_2\!=\!-0.023$ at physical quark mass in full QCD, we obtain 
\bea
   f_+(0) = 0.968(9)(6),
   \label{eqn:chiral_fit:f+}
\eea
which is consistent with the previous lattice calculations
\cite{kl3:fn:Nf0:italy,kl3:fn:Nf2:JLQCD,kl3:fn:Nf3:FNAL},
employing different discretizations (Wilson and KS)
in quenched and unquenched QCD,
listed in Table~\ref{tbl:chiral_fit:other},
as well as with estimates based on $O(p^6)$ ChPT
\cite{kl3:f4:ChPT:BT,kl3:f4:ChPT:JOP,kl3:f4:ChPT:CEEKPP},
and the Leutwyler-Roos's value \cite{kl3:f2+f4:model:LR}.


In Eq.~(\ref{eqn:chiral_fit:f+}), 
we have not included systematic uncertainties due to the discretization 
error and effects of dynamical strange quarks, which are difficult to 
estimate reliably without simulations at different lattice spacings
or those in three-flavor QCD.
However, these uncertainties affect our estimate of $f_+(0)$ only through 
the small higher order correction $\Delta f$, and hence are expected 
not to be large.
This is supported by the nice consistency with results from different lattice 
actions and/or with different numbers of flavors for dynamical quarks
(see Table~\ref{tbl:chiral_fit:other}).
We also note that the RBC and UKQCD Collaborations have already started 
large-scale simulations with three flavors of dynamical domain-wall quarks,
and their preliminary estimate 
is consistent with Eq.~(\ref{eqn:chiral_fit:f+}).
In particular, that comparison should provide a more reliable estimate
of systematic error due to quenching of strange quarks.

\begin{figure}[b]
\vspace{2mm}
\begin{center}

\includegraphics[width=0.5\linewidth,clip]{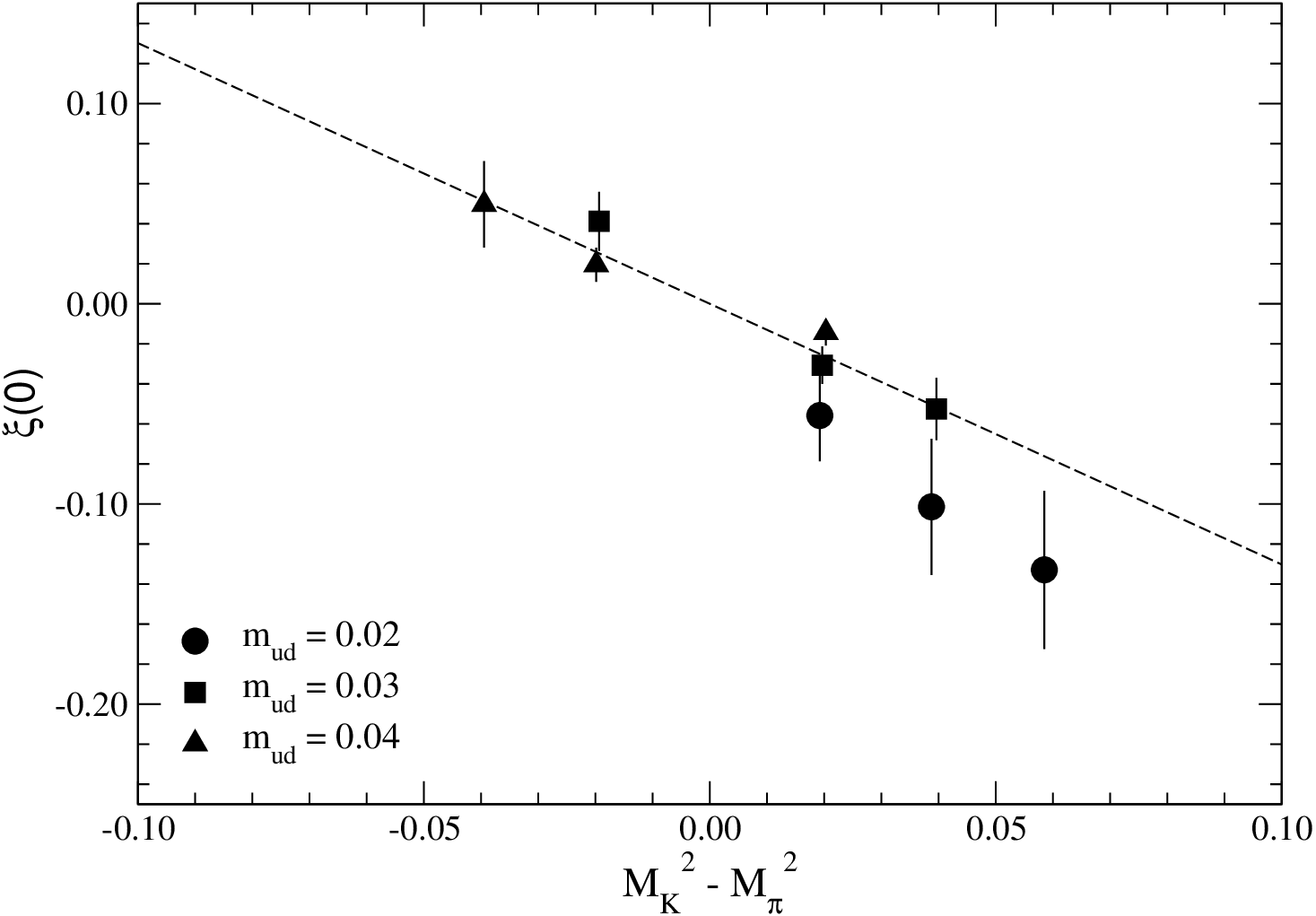}

\end{center}
\vspace{0mm}
\caption{
   Chiral extrapolation of $\xi(q^2)$.
}
\label{fig:chiral_fit:xi}
\end{figure}

We also calculate $\xi(0)$ at the physical quark mass.
Since $\xi(0)$ vanishes in the $SU(3)$ symmetric limit, 
we test the following simple linear fit 
\bea
   \xi(0) = d_{1,v} (M_K^2-M_{\pi}^2),
   \label{eqn:chiral_fit:xi:form:val-lin}
\eea
and find that this leads to a reasonable value of $\chi^2/{\rm dof}$
as shown in Table~\ref{tbl:chiral_fit:xi}.
The fit line is plotted in Fig.~\ref{fig:chiral_fit:xi}.
We obtain 
\bea
   \xi(0) = -0.105(22),
   \label{eqn:chiral_fit:xi:result}
\eea
which is consistent with the experimental values 
$-0.01(6)$ for $K_{l3}^0$ and $-0.125(23)$ for $K_{l3}^+$ \cite{PDG2004}.


\section{$|V_{us}|$ and CKM unitarity}
\label{sec:Vus}

By combining with an estimate of $|V_{us}\,f_+(0)|\!=\!0.2173(8)$ 
based on the recent experimental determination of $\Gamma$ \cite{CKM2005}, 
we obtain 
\bea
   |V_{us}| = 0.2245(26)(8),
   \label{eqn:Vus:V_us}
\eea
where the first and second errors come from the uncertainty in $f_+(0)$ 
and $|V_{us}\,f_+(0)|$, respectively.
This leads to 
\bea
   |V_{ud}|^2+|V_{us}|^2+ |V_{ub}|^2 = 1 - \delta, \hspace{5mm}
   \delta = 0.0013(16),
   \label{eqn:Vus:unitarity}
\eea
which is completely consistent with CKM unitarity.


\section{Conclusion}
\label{sec:conclusion}


In this paper, 
we have calculated $f_+(0)$ from numerical simulations of two-flavor 
dynamical QCD using the domain-wall quark action.
We obtained 
\bea
   f_+(0) = 0.968(9)(6), \hspace{5mm}
   |V_{us}| = 0.2245(26)(8),
\eea
which supports CKM unitarity.
While we have not estimated systematic uncertainties due to the 
use of the finite lattice spacing and the quenched approximation for 
strange quarks, these are expected to be small from the nice consistency
with other lattice estimates.


Our result for $f_+(0)$ is consistent with the phenomenological 
estimate, which has been used in previous determinations of $|V_{us}|$,
and hence has not changed $|V_{us}|$ significantly.
The main significance of this study is that now 
$f_+(0)$ has been calculated non-perturbatively from two-flavor QCD 
and its uncertainties can be systematically reduced in future lattice 
calculations.
Systematic errors which all the present lattice calculations share are those
connected with the interpolation in momenta and extrapolation in mass. 
In both these cases it was necessary to use an ansatz. 
Since the interpolation in the momentum transfer was over a very small range,
and the extrapolation in mass systematically took into account the calculated
behavior up to NLO in ChPT, relying on the ansatz only for higher orders,
both these effects are expected to be small. 

However, it should be noted that the sea quark masses used in this calculation
are relatively heavy, and to be confident that ChPT is a good description of 
the data it would be advisable to move to smaller masses.
In turn this will make the momentum interpolation more difficult 
as $q_{\rm max}^2$ deviates further from 0.
Another important step in the future is clearly an extension to 
dynamical three-flavor QCD. 
The RBC and UKQCD Collaborations' study of three-flavor QCD is well underway
\cite{ConfGene:Nf3QCD:RBC-UKQCD,Kl3:Nf3QCD:RBC-UKQCD},
and a more reliable estimate of systematic uncertainties in $f_+(0)$ 
will come in the near future.

\begin{acknowledgments}

The authors would like to thank members of our RBC Collaboration and 
especially Norman Christ for many discussions and support.
We also thank RIKEN, Brookhaven National Laboratory (BNL) 
and the U.S. Department of Energy for providing
the facilities essential for this work.
The work of TK is supported in part by the Grant-in-Aid of the
Japanese Ministry of Education (Nos.17740171).
The work of AS was supported in part by US DOE Contract No.
DE-AC02-98CH10886.
TK is grateful to the Theory Group in BNL for their kind hospitality 
during his stay when this work was initiated.

\end{acknowledgments}


\clearpage


\begin{table}[htbp]
\caption{
   Fitted meson masses.
}
\begin{center}
\begin{ruledtabular}
\begin{tabular}{lll|lll|lll}
   \multicolumn{2}{l}{$m_{ud}$} & \multicolumn{1}{l|}{$M_{\pi}$} &
   \multicolumn{2}{l}{$m_{ud}$} & \multicolumn{1}{l|}{$M_{\pi}$} &
   \multicolumn{2}{l}{$m_{ud}$} & \multicolumn{1}{l}{$M_{\pi}$} 
   \\ \hline
   \multicolumn{2}{l}{0.02}     & \multicolumn{1}{l|}{0.2927(18)} & 
   \multicolumn{2}{l}{0.03}     & \multicolumn{1}{l|}{0.3570(26)} & 
   \multicolumn{2}{l}{0.04}     & \multicolumn{1}{l}{0.4081(21)} 
   \\ \hline 
   $m_{ud}$  & $m_s$  & $M_{K}$ & 
   $m_{ud}$  & $m_s$  & $M_{K}$ & 
   $m_{ud}$  & $m_s$  & $M_{K}$ 
   \\ \hline
   0.02      & 0.03   & 0.3239(17) & 
   0.03      & 0.02   & 0.3288(28) & 
   0.04      & 0.02   & 0.3565(22) 
   \\
   0.02      & 0.04   & 0.3528(17) &
   0.03      & 0.04   & 0.3836(24) &
   0.04      & 0.03   & 0.3830(20) 
   \\
   0.02      & 0.05   & 0.3797(17) &
   0.03      & 0.05   & 0.4088(24) &
   0.04      & 0.05   & 0.4323(21) 
\end{tabular}       
\end{ruledtabular}
\end{center}
\label{tbl:ff:mass:mass}
\end{table}


\begin{table}[htbp]
\caption{
   Scalar form factor $f_0(q_{\rm max}^2)$ at simulated quark masses.
}
\begin{center}
\begin{ruledtabular}
\begin{tabular}{lll|lll|lll}
   $m_{ud}$ & $m_s$ & $f_0(q_{\rm max}^2)$ &
   $m_{ud}$ & $m_s$ & $f_0(q_{\rm max}^2)$ &
   $m_{ud}$ & $m_s$ & $f_0(q_{\rm max}^2)$ 
   \\ \hline
   0.02     & 0.03   & 1.00067(17) & 
   0.03     & 0.02   & 1.00050(22) &
   0.04     & 0.02   & 1.00098(55) 
   \\
   0.02     & 0.04   & 1.00202(48) &
   0.03     & 0.04   & 1.00036(11) &
   0.04     & 0.03   & 1.00024(10) 
   \\
   0.02     & 0.05   & 1.00352(82) &
   0.03     & 0.05   & 1.00126(35) &
   0.04     & 0.05   & 1.00018(6)
\end{tabular}       
\end{ruledtabular}
\end{center}
\label{tbl:ff:f0:f0}
\end{table}


\begin{table}[htbp]
\caption{
   Fit parameters for interpolation of $f_0(q^2)$ to $q^2\!=\!0$ 
   using pole form Eq~(\ref{eqn:q2_interp:form:pole}).
}
\begin{center}
\begin{ruledtabular}
\begin{tabular}{lll|lll|lll}
   $m_{ud}$  & $ m_s$  & $\chi^2/\mbox{dof}$  & $f_0(0)$  & $\lambda_0^{(1)}$
   \\ \hline
   0.02  & 0.03  & 0.76  & 0.99955(47)  & 1.16(39)  \\
   0.02  & 0.04  & 0.67  & 0.9979(14)   & 1.13(32)  \\
   0.02  & 0.05  & 0.60  & 0.9952(25)   & 1.09(28)  \\
   \hline
   0.03  & 0.02  & 1.49  & 0.99872(43)  & 2.23(40)  \\
   0.03  & 0.04  & 1.91  & 0.99912(26)  & 1.75(32)  \\
   0.03  & 0.05  & 2.07  & 0.99697(90)  & 1.60(30)  \\
   \hline
   0.04  & 0.02  & 1.98  & 0.99574(95)  & 1.97(33)  \\
   0.04  & 0.03  & 1.79  & 0.99911(20)  & 1.80(28)  \\
   0.04  & 0.05  & 1.06  & 0.99924(15)  & 1.61(23)  \\
\end{tabular}       
\end{ruledtabular}
\end{center}
\label{tbl:q2_interp:mthd1}
\end{table}

\begin{table}[htbp]
\caption{
   Results for $\xi(0)$ and $f_0(0)$ obtained from
   Eqs.~(\ref{eqn:q2_interp:F24_vs_q2})\,--\,(\ref{eqn:q2_interp:xi_vs_q2}).
}
\begin{center}
\begin{ruledtabular}
\begin{tabular}{ll|ll}
   $m_{ud}$  & $ m_s$  & $\xi(0)$  & $f_0(0)$  
   \\ \hline
   0.02      & 0.03  & -0.056(23)  & 1.00007(94)  \\
   0.02      & 0.04  & -0.101(34)  & 0.9992(29)   \\
   0.02      & 0.05  & -0.133(39)  & 0.9962(57)   \\
   \hline
   0.03      & 0.02  & +0.041(15)  & 1.00017(58)  \\
   0.03      & 0.04  & -0.0307(94) & 0.99985(36)  \\
   0.03      & 0.05  & -0.053(16)  & 0.9990(12)   \\
   \hline
   0.04      & 0.02  & +0.050(22)  & 0.9968(15)   \\
   0.04      & 0.03  & +0.0195(85) & 0.99915(31)  \\
   0.04      & 0.05  & -0.0144(65) & 0.99912(24)  \\
\end{tabular}       
\end{ruledtabular}
\end{center}
\label{tbl:q2_interp:mthd2}
\end{table}


\begin{table}[htbp]
\caption{
   Fit parameters for chiral extrapolation of $f_+(0)$,
   and $\Delta f$ at physical quark mass.
}
\begin{center}
\begin{ruledtabular}
\begin{tabular}{l|llll|l}

   fit form  & $\chi^2/\mbox{dof}$ & $c_0$ & $c_{1,s}$ & $c_{1,v}$ & $\Delta f$
   \\ \hline
   Eq.~(\ref{eqn:chiral_fit:fit_form:const})  & 
   0.46 & -1.99(34) & --        & --        & -0.013(2)  \\
   Eq.~(\ref{eqn:chiral_fit:fit_form:val-lin})  & 
   0.46 & -1.2(2.0) & --        & -2.6(6.3) & -0.009(9)  \\
   Eq.~(\ref{eqn:chiral_fit:fit_form:lin})  & 
   0.10 & -2.3(1.8) & -44(14)   & 22.1(5.4) & -0.003(11) \\
\end{tabular}       
\end{ruledtabular}
\end{center}
\label{tbl:chiral_fit:f+}
\end{table}

\begin{table}[htbp]
\caption{
   Recent lattice estimates of $f_+(0)$. Note that unquenched results 
   in Ref.\cite{kl3:fn:Nf2:JLQCD,kl3:fn:Nf3:FNAL} are preliminary.
   Two values in Ref.\cite{kl3:fn:Nf2:JLQCD} are obtained 
   from two choices of the chiral extrapolation form 
   (polynomial and ChPT based forms).
}
\begin{center}
\begin{ruledtabular}
\begin{tabular}{l|ll|l}
   & $N_f$ & quark action & $f_+(0)$
   \\ \hline
   this work  
   & 2 & domain-wall        & 0.968(11)
   \\ \hline 
   Becirevic~{\it et al.}~\cite{kl3:fn:Nf0:italy}
   & 0 & improved Wilson    & 0.960(9) 
   \\
   JLQCD~\cite{kl3:fn:Nf2:JLQCD}
   & 2 & improved Wilson    & 0.967(6), 0.952(6)
   \\
   MILC~\cite{kl3:fn:Nf3:FNAL}
   & 3 & improved staggered & 0.962(11) 
\end{tabular}       
\end{ruledtabular}
\end{center}
\label{tbl:chiral_fit:other}
\end{table}

\begin{table}[htbp]
\caption{
   Fit parameter for chiral extrapolation of $\xi(0)$.
}
\begin{center}
\begin{ruledtabular}
\begin{tabular}{l|llll|l}
   $\chi^2/\mbox{dof}$ & $d_{1,v}$
   \\ \hline
   0.45                & -1.30(28) 
\end{tabular}       
\end{ruledtabular}
\end{center}
\label{tbl:chiral_fit:xi}
\end{table}


\clearpage



\end{document}